\documentclass[12pt,preprint]{aastex}

\def\lesssim{\mathrel{\hbox{\rlap{\hbox{\lower4pt\hbox{$\sim$}}}\hbox{$<$}}}}
\def\gtrsim{\mathrel{\hbox{\rlap{\hbox{\lower4pt\hbox{$\sim$}}}\hbox{$>$}}}}
\providecommand{\etal}{et~al.}

\def\gax{\mathrel{\raise.3ex\hbox{$>$}\mkern-14mu\lower0.6ex\hbox{$\sim$}}}
\def\lax{\mathrel{\raise.3ex\hbox{$<$}\mkern-14mu\lower0.6ex\hbox{$\sim$}}}
\def\gtorder{\mathrel{\raise.3ex\hbox{$>$}\mkern-14mu
             \lower0.6ex\hbox{$\sim$}}}
\def\ltorder{\mathrel{\raise.3ex\hbox{$<$}\mkern-14mu
             \lower0.6ex\hbox{$\sim$}}}

\begin{document}

\title{The Unusual Temporal and Spectral Evolution 
of the Type IIn Supernova 2011ht\altaffilmark{1}}

\author{P.~W.~A. Roming\altaffilmark{2,3}, 
T.~A. Pritchard\altaffilmark{3}, J.~L. Prieto\altaffilmark{4,5}, 
C.~S. Kochanek\altaffilmark{6,7}, C.~L. Fryer\altaffilmark{8,9,10}, 
K. Davidson\altaffilmark{11}, R.~M. Humphreys\altaffilmark{11}, 
A.~J. Bayless\altaffilmark{2}, J.~F. Beacom\altaffilmark{6,7,12}, 
P.~J. Brown\altaffilmark{13}, S.~T. Holland\altaffilmark{14}, 
S. Immler\altaffilmark{15,16,17}, 
N.~P.~M. Kuin\altaffilmark{18}, S.~R. Oates\altaffilmark{18}, 
R.~W. Pogge\altaffilmark{6,7}, G. Pojmanski\altaffilmark{19}, 
R. Stoll\altaffilmark{6}, B.~J. Shappee\altaffilmark{6,20}, 
K.~Z. Stanek\altaffilmark{6,7}, D.~M. Szczygiel\altaffilmark{6}} 

\altaffiltext{1}{This paper is dedicated to our colleague, 
Weidong Li, who died on 2011-12-12. His contribution
to the study of all types of SNe was tremendous.}

\altaffiltext{2}{Space Science \& Engineering Division,
Southwest Research Institute, P.O. Drawer 28510, San Antonio, TX
78228-0510, USA; Corresponding author's e-mail: proming@swri.edu}

\altaffiltext{3}{Department of Astronomy \& Astrophysics,
Penn State University, 525 Davey Lab, University Park, PA 
16802, USA}

\altaffiltext{4}{Department of Astrophysical Sciences, Princeton
University, Peyton Hall, Princeton, NJ 08544, USA}

\altaffiltext{5}{Hubble and Carnegie-Princeton Fellow}

\altaffiltext{6}{Department of Astronomy, The Ohio State 
University, 140 W. 18th Ave., Columbus, OH 43210, USA}

\altaffiltext{7}{Center for Cosmology \& AstroParticle Physics, 
The Ohio State University, 191 W. Woodruff Ave., Columbus, 
OH 43210, USA}

\altaffiltext{8}{CCS-2, MS D409, Los Alamos National 
Laboratory, Los Alamos, NM, 87544, USA}

\altaffiltext{9}{Physics Department, University of Arizona, 
Tucson, AZ 85721, USA}

\altaffiltext{10}{Physics and Astronomy Department, University 
of New Mexico, Albuquerque, NM 87131, USA}

\altaffiltext{11}{Minnesota Institute for Astrophysics, 
University of Minnesota, 116 Church St. S.E., Minneapolis, 
MN 55455, USA}

\altaffiltext{12}{Department of Physics, The Ohio State University, 
191 W. Woodruff Ave., Columbus, OH 43210, USA}

\altaffiltext{13}{Department of Physics \& Astronomy, 
University of Utah, 201 James Fletcher Bldg., 
115 S. 1400 E. 201, Salt Lake City, UT 84112-0830, USA}

\altaffiltext{14}{Space Telescope Science Center, 3700
San Martin Dr., Baltimore, MD 21218, USA}

\altaffiltext{15}{Astrophysics Science Division, NASA Goddard 
Space Flight Center, Greenbelt, MD 20771, USA}

\altaffiltext{16}{Center for Research \& Exploration in Space 
Science \& Technology, NASA Goddard Space Flight Center, 
Greenbelt, MD 20771, USA}

\altaffiltext{17}{Department of Astronomy, University of 
Maryland, College Park, MD 20742, USA}

\altaffiltext{18}{Mullard Space Science Laboratory, University 
College London, Holmbury St. Mary, Dorking, Surrey RH5 6NT, UK}

\altaffiltext{19}{Warsaw University Astronomical Observatory, 
Al. Ujazdowskie 4, 00-478 Warsaw, Poland}

\altaffiltext{20}{National Science Foundation Fellow}

\begin{abstract}
We present very early UV to optical photometric and spectroscopic 
observations of the peculiar Type IIn supernova (SN) 2011ht in 
UGC~5460. The UV observations of the rise to peak are only the 
second ever recorded for a Type IIn SN and are by far the most 
complete. The SN, first classified as a 
SN impostor, slowly rose to a peak of $M_V\sim-17$ in $\sim$55\, 
days. In contrast to the $\sim$2 magnitude increase in 
the $v$-band light curve from the first observation until 
peak, the UV flux increased by $>$7 magnitudes. The optical spectra 
are dominated by strong, Balmer emission with narrow peaks 
(FWHM$\sim$600\,km\,s$^{-1}$), very broad asymmetric wings 
(FWHM$\sim$4200\,km\,s$^{-1}$), and blue shifted 
absorption ($\sim$300\,km\,s$^{-1}$) superposed on a strong 
blue continuum. The UV spectra are dominated by 
\ion{Fe}{2}, \ion{Mg}{2}, \ion{Si}{2}, and \ion{Si}{3} 
absorption lines broadened by $\sim$1500\,km\,s$^{-1}$. 
Merged X-ray observations reveal a 
$L_{0.2-10} = (1.0\pm0.2)\times10^{39}$\,erg\,s$^{-1}$. 
Some properties of SN~2011ht are similar to SN impostors,
while others are comparable to Type IIn SNe. Early spectra
showed features typical of luminous blue
variables at maximum and during giant 
eruptions. However, the broad 
emission profiles coupled with the strong UV flux have not been 
observed in previous SN impostors. The absolute 
magnitude and energetics ($\sim2.5\times10^{49}$\,~ergs in the 
first 112 days) are reminiscent of normal Type IIn SN, 
but the spectra are of a dense wind. 
We suggest that the mechanism for creating 
this unusual profile could be a shock interacting with a shell of material 
that was ejected a year before the discovery of the SN.
\end{abstract}


\section {Introduction}
Type IIn supernovae (SNe~IIn), a designation first 
proposed by \citet{sem90}, are characterized by spectra with 
a narrow H$\alpha$ emission line superimposed on a broad
emission component \citep{cp09} that lacks a broad P-Cygni
absorption feature \citep{pa02}. Most SNe~IIn are core collapse 
events that are due to shock interaction of the ejecta with the 
circumstellar material \citep[CSM;][]{cnn90,pa02,sn08a,cp09},
although some thermonuclear explosions are thought to detonate in
a hydrogen-rich environment producing SNe~IIn-like events 
\citep[cf.][]{hm03,pa05,pjl07}. As the shock interacts and heats 
the dense CSM, there is a rapid increase in the production of
UV photons. 

SNe~IIn are a very heterogeneous group 
\citep[e.g.][]{cp09,oeo10,km10} ranging 
from the most luminous SNe 
\citep[e.g. SN~2006gy with $M_V \simeq -22$ and 
SN~2008am with $M_R \simeq -22.3$;][]{sn07,ce11} 
to some of the fainter
\citep[e.g. SNe~1997bs and 2008S with $M_V \simeq -13$ and 
$M_R \simeq -13.9$, respectively;][]{vds00,sn09}. 
The nature of the fainter objects, such as 
SNe~1997bs and 2008S, is unclear \citep{vds00,sn09}. 
Many argue that they 
are non-terminal outbursts as opposed to
core collapse objects \citep[e.g.][]{sn09} 
and have been designated
``supernova impostors" \citep{vds00}. 
The great eruption of $\eta$~Car is frequently 
cited as a Galactic analogue. However, this may be 
misleading since the great eruption of 
$\eta$~Car was an unusual 
event and none of the extragalactic impostors
resemble it in terms of kinetic energy 
and duration \citep[see][]{sn11,kcs11b,kcs12}. Although some
impostors are clearly non-terminal events (e.g. $\eta$~Car,
P-Cygni, SN~2000ch), it is not clear that this is the
case for all fainter outbursts 
\citep[cf.][but see Humphreys \etal 2011]{pjl08,bmt09,tta09,kcs11}.
For the brighter SNe~IIn, there is mounting 
evidence that the progenitor is a
massive, luminous star 
that retains its hydrogen-shell until
shortly before explosion and that the terminal 
mass loss event is strongly correlated with 
the death of the star \citep{gya07,sn11,kcs11b}. 

With the emergence of dedicated supernovae (SNe) follow-up 
programs and telescopes \citep[e.g.][]{fav01,bjs06,hm06,qrm06,
daj09,lnm09,ra09}, SNe~IIn are now being studied more frequently 
in both the optical and near-IR 
\citep[e.g.][]{pa02,pjl08,sn08a,sn09,daj10,km10,sn10,daj11,sn11a}. 
UV studies, 
on the other-hand, have seriously lagged behind the redder 
wavelengths, even though the UV is a promising probe of these 
interesting objects. Unlike other classes of SNe, SNe~IIn are 
very bright in the UV and can be found in $z>2$ optical surveys 
while Type~Ia SNe have much redder spectra \citep{pn03,cj09}.
Their intrinsic brightness and long-lived emission also 
aid in spectroscopically confirming such distant 
SNe \citep{cj08}. Since they are associated with massive 
stars and because of their intrinsic UV brightness, 
SNe~IIn are potentially strong probes of star formation and galaxy
evolution out to relatively large redshifts \citep[$z>2$;][]{cj08}. 

Previous to 2005, only three SNe~IIn had been observed in the 
UV \citep[SNe~1988Z, 1994Y, and 1998S;][]{bpj09}, 
and only a handful since\footnote{see 
http://swift.gsfc.nasa.gov/docs/swift/sne/swift\_sn.html}. However, none of
these SNe were observed early enough to detect the rise in the
UV light curve (despite the rapid follow-up after discovery 
in the case of {\em Swift} UVOT observed SNe~IIn). 
Recently, the SN~IIn PTF09uj was caught on the 
rise \citep[$\sim$1.8\,mag in $\sim$4\,days based
on two UV detections;][]{oeo10}.
This rapid UV rise was attributed to the supernova (SN) shock breaking
through a dense CSM. 

In this paper, we describe SN~2011ht, where the rise in the UV
was captured in far greater detail, with over 10 UV observations
before peak. Unlike PTF09uj, the rise to peak for SN~2011ht was
large ($\sim$7\,mag) and slow ($\sim$40\,days).
In Section 2, we present the UV to optical photometric and spectroscopic 
observations, X-ray observations, pre-explosion images, and the 
host galaxy properties.
In Section 3, we review the UV/optical light curves, blackbody fits, 
and spectral properties. In Section 4, we discuss our results including
a comparison to other SN impostors and SNe~IIn, as well as model
interpretations. 
In Section 5, we summarize our conclusions.

\section{Observations}
\subsection{Discovery}
\begin{figure} 
\epsscale{0.9} 
\plotone{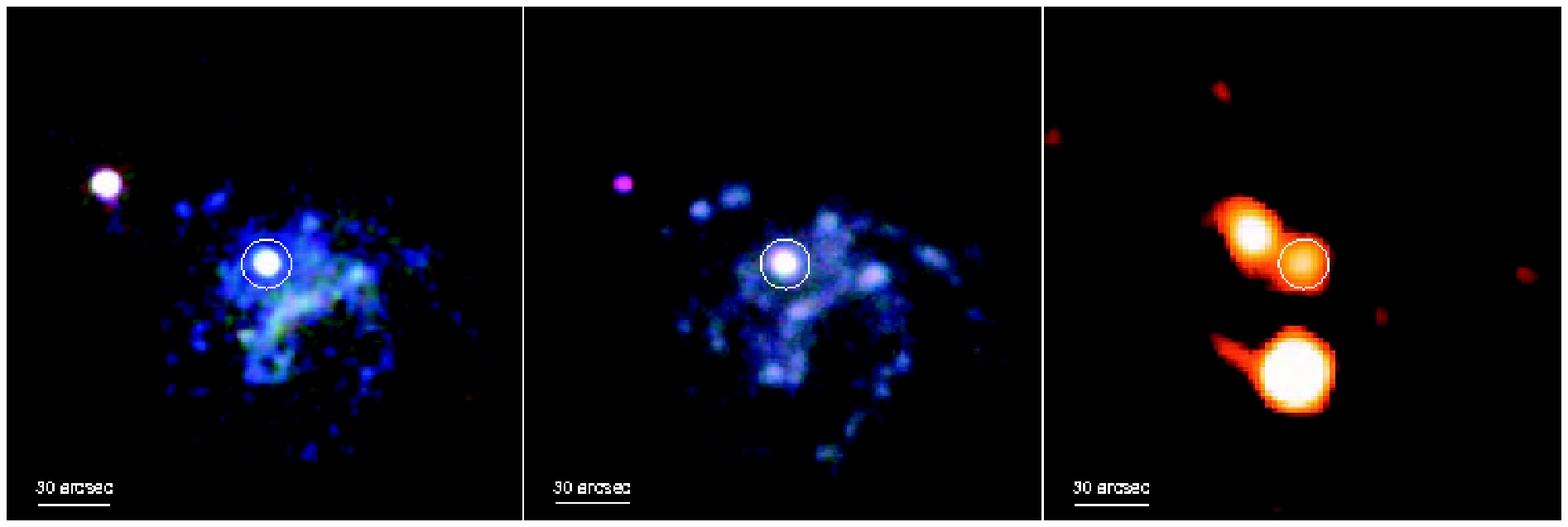} 
\caption{{\em Swift} UVOT optical ({\em left}), UVOT UV 
({\em middle}), and XRT X-ray ({\em right}) images of SN~2011ht 
and its host galaxy UGC~5460. {\em Left:} The optical image was 
constructed from the UVOT $v$ (64\,s; red), $b$ (64\,s; green), 
and $u$ (64\,s; blue) filters obtained on 2011-11-02. {\em Middle:} 
The UV image was constructed from the UVOT $uvw1$ (20\,s; red), 
$uvm2$ (20\,s; green), and $uvw2$ (33\,s; blue) filters obtained 
on 2011-11-02. {\em Right:} The X-ray image (0.2--10\,keV) was 
constructed from merged XRT data obtained 
between 2011-10-02 and 2012-01-17 (88.1\,ks exposure time). 
All images are smoothed with a 
3\,pixels FWHM Gaussian. The position of SN~2011ht 
is indicated by a circle with a radius of $10\arcsec$.} 
\label{fig-2011ht}    
\end{figure}

SN~2011ht (Figure~\ref{fig-2011ht}) was discovered on 2011-09-29.182  
(UT; JD~2455833.682) in the nearby galaxy ($z=0.0036$) UGC~5460 
at (J2000) ${\rm R.A.} = 10^{h}08^{m}10^{s}.58$, 
${\rm Dec.} = +51^{\circ} 50\arcmin 57\farcs5$ with a 
magnitude of 16.9 \citep{bt11}. An ALFOSC spectrum taken with the Nordic Optical 
Telescope on 2011-09-30.22 suggested that the object
was a ``SN impostor" with properties similar to the ``giant eruption" \citep[see][]{hd94}
of the luminous blue variable UGC~2773-OT \citep{pa11}. The object was designated
PSN~J10081059+5150570. On 2011-10-18, \citet{rp11} reported 
UV ($uvw2$, $uvm2$, and $uvw1$), $u$, $b$, and $v$-band
brightenings of $\sim$4, $\sim$3, $\sim$1.5 and $\sim$1.5 magnitudes, 
respectively. The transient also brightened another
0.4\,mag between 2011-10-22 and 2011-10-29, to $V\sim 14.7$.  
At a distance of 19.2\,Mpc for the host galaxy UGC~5460 \citep{pa11}, this corresponds  
to an absolute magnitude of $M_V \sim -17$ that is typical of
a core-collapse SN \citep{lc11}. Based in part on this high luminosity, \citet{pjl11} 
suggested that PSN~J10081059+5150570 was a SNe~IIn
rather than an impostor. Spectroscopic observations 
on 2011-11-11.5 revealed that the emission profiles were 
similar to some SNe~IIn \citep{pjl11b}. 
 
\subsection{Photometric Data}
\begin{figure} 
\epsscale{0.9} 
\plotone{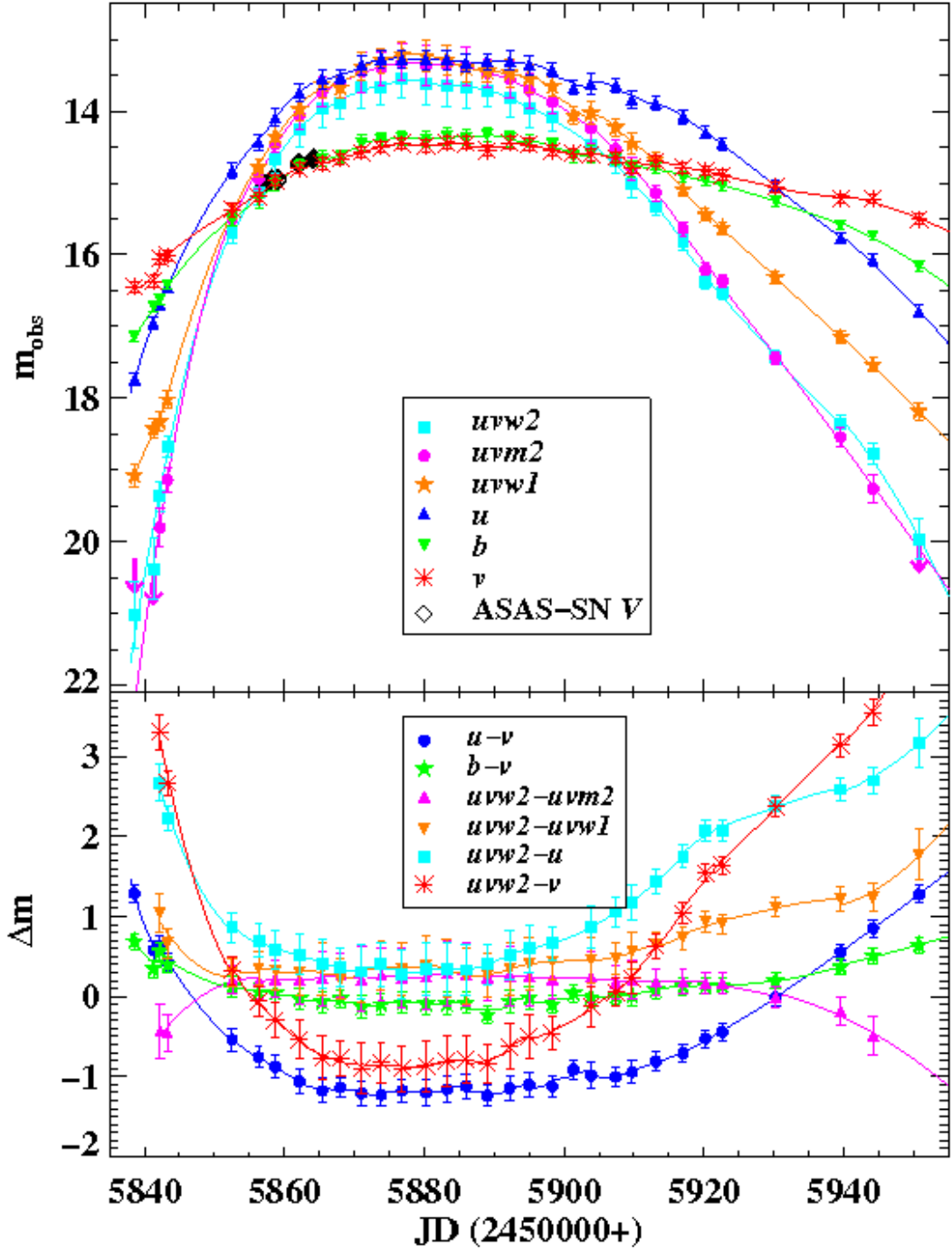} 
\caption{{\em Top: Swift} UVOT and ASAS-SN light curves of SN~2011ht.
{\em Bottom:} Color evolution of SN~2011ht.} 
\label{fig-lc}    
\end{figure}

Photometric observations were obtained with the {\em Swift} 
\citep{gn04} Ultra-Violet/OpticalTelescope \citep[UVOT;][]{rp05,rp04,rp00}
and the All Sky Automated Survey (ASAS-SN) North telescope
located on Haleakala, Hawaii. 
The photometric observations with the UVOT began on 2011-10-4.9, 
5.8 days after the initial discovery. The observations were
obtained for the three optical filters ($ubv$) and the three UV filters 
\citep[$uvw2$, $uvm2$, $uvw1$: $\lambda_c = 1928, 2246, 2600$\,\AA, 
respectively;][]{pt08} with a median cadence of three days. Photometry 
was performed using a $3\arcsec$ 
source aperture following the method outlined in \citet{bpj09b}. 
A $20\arcsec$ aperture centered on a region in the galaxy with a similar
background to SN~2011ht was used for galaxy subtraction.
The data reduction pipeline used the HEASOFT~6.6.3 
(which includes time-dependent sensitivity corrections)
and {\em Swift} Release~3.3 analysis tools with 
updated UVOT zero-points from \citet{baa10} to calibrate the 
resulting photometry (Table~\ref{tab1}). The UVOT light curves
and the corresponding color evolution are shown in Figure~\ref{fig-lc}.
We note that there is a +0.076 ($\sigma = 0.052$), +0.010 
($\sigma = 0.049$), and 0.068 ($\sigma = 0.027$) magnitude 
($3\sigma$ confidence limit) offset between the UVOT and SDSS $u$, $b$, and $v$ filters, 
respectively \citep{rp09}.
 
We also obtained $V$-band photometric observations (Figure~\ref{fig-lc}) 
of SN~2011ht using the commissioning data from the recently  
installed, $2\times$14-cm diameter ASAS-SN telescope.  
We used ASAS-SN Unit1, equipped with a FLI ProLine CCD camera with  
Fairchild Imaging $2{\rm k}\times 2{\rm k}$ thinned CCD, giving a $4.47 \times  
4.47\;$square-degree field-of-view, corresponding to $7.8\arcsec/{\rm pixel}$ image  
scale. Because such a large image scale leads to a significant  
flux contamination by the host galaxy, we estimated the amount of  
contamination using the {\em Swift} satellite data and used this to
set the baseline of the ASAS-SN photometry. 

\subsection{Spectroscopic Data}
\begin{figure} 
\centering
\includegraphics[width=0.75\textwidth,angle=90]{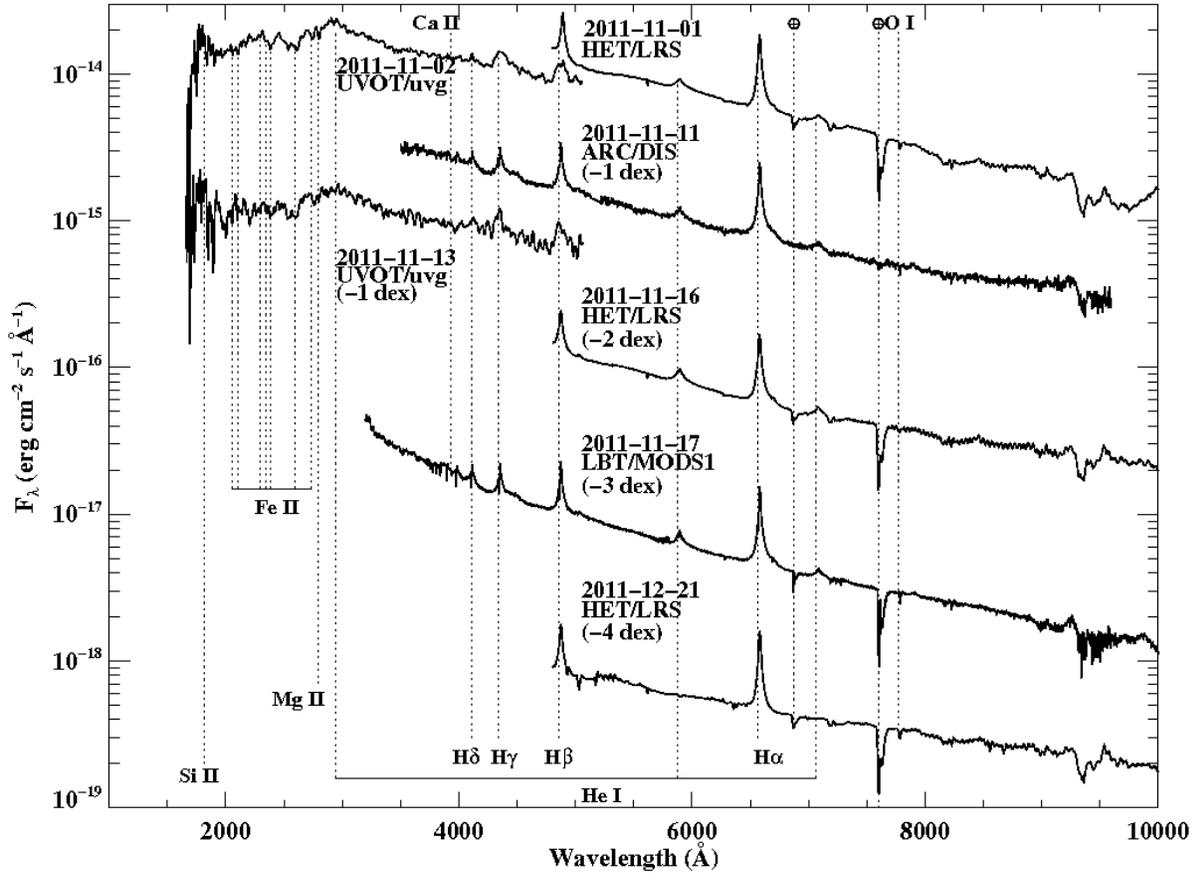} 
\caption{HET/LRS, UVOT/uvg, ARC/DIS, and LBT/MODS1 spectra of SN~2011ht. 
Prominent absorption and emission lines, as well as 
telluric bands, are indicated. The wavelengths have 
been corrected to the rest frame using $v_{\rm {helio}} =
1093$\,km\,s$^{-1}$. The spectra have not been corrected for 
line-of-sight extinction.} 
\label{fig-spect}    
\end{figure}

Spectroscopic observations were obtained with 
the Hobby Eberly Telescope \citep[HET;][]{rl98}
Low Resolution Spectrograph \citep[LRS;][]{hgj98}, 
{\em Swift} UVOT UV grism (uvg), Astrophysical 
Research Consortium (ARC) telescope Dual Imaging Spectrograph 
(DIS), and Large Binocular Telescope (LBT) Multi-Object Double 
Spectrograph \#1 \citep[MODS1;][]{prw10}. Dates
and exposure times for each of the instruments
are provided in Table~\ref{tab2} and the
corresponding spectra are shown in Figure~\ref{fig-spect}.

The HET spectra were obtained on 2011-11-01, 2011-11-16, and 2011-12-21. 
The HET/LRS was used with a $2\arcsec$ slit (${\rm R}\approx 300$; 
$\Delta\lambda \approx 4,500 - 10,000$\,\AA). Standard IRAF\footnote{IRAF is 
distributed by the National Optical Astronomy Observatory, which is operated by the 
Association of Universities for Research in Astronomy (AURA) under cooperative 
agreement with the National Science Foundation.} reduction techniques of bias 
subtraction, flat fielding, and wavelength calibration were used. Relative flux 
calibration was performed using the flux standards GD191B2B and HD84937 \citep{fm96,mg90}. 

The {\em Swift} UVOT observations of SN~2011ht included 
grism observations. On 2011-11-02 and 2011-11-13, four exposures 
for a total of 3710\,s and thirteen exposures for 
a total of 15,551\,s were completed, respectively. 
The data for the individual exposures 
were extracted using the new UVOT grism wavelength calibration (Kuin \etal, 
in prep) and the current flux calibration from the {\em Swift} 
UVOT CALDB (version 20110731). Based on this calibration, 
the accuracy of the wavelength scale 
is typically 15\,\AA. The individual exposures were 
summed after rotating the images along the dispersion direction 
and alignment of the anchor positions. The anchor positions were 
defined by the 2600\,\AA\ line observed in first order. Taking into account the 
curvature of the spectrum, the spectrum was then 
extracted from the summed image using a slit width of $1\sigma$ of 
the cross-dispersion Gaussian. 

The errors in the flux due to noise are low for wavelengths
$\gtrsim$1900\,\AA, but the signal-to-noise (S/N) is $<$3 below 
1910\,\AA\ and 1770\,\AA\ (which is approximately where the 
sensitivity in the grism begins to drop off) for the 
2011-11-02 and 2011-11-13 spectra, respectively. The position
of the spectra on the detector was such 
that a large portion of the second order spectrum
lies next to the first order spectrum at wavelengths less then 
3600\,\AA\ and partially overlaps at 4550\,\AA. 
At this point the contaminating flux 
due to the second order is estimated to be about $12\%$
of the total flux.

We also obtained a spectrum of SN~2011ht on 2011-11-11.5 
with DIS on the 3.5m ARC telescope at the  
Apache Point Observatory (APO). 
The observations were done with the B400/R300 low-resolution gratings and  
a $1\farcs5$ slit, which give a wavelength coverage  
from $\sim 3500-9600$\,\AA\ with a FWHM resolution  
of 7\,\AA. Standard IRAF tasks were used to reduce the 
spectrum. The data were calibrated using a HeNeAr
lamp and standard star observations taken on the
same night. The strongest telluric lines were corrected 
using the spectrophotometric standards.

On 2011-11-17.5 spectra were obtained with 
the MODS1 instrument using a $1\farcs0$ wide 
slit and the G400L and G750L gratings in the 
Blue and Red channels, respectively. 
The observations were done in a sequence of 
three 300\,s integrations for a total integration 
time of 900\,s. The slit was oriented along 
the mean parallactic angle during the exposures 
($PA=-140\deg$) since the MODS1 
spectrograph does not have an atmospheric 
dispersion corrector (ADC).  
The seeing during the exposures was 
$0\farcs6$ FWHM, giving a spectral resolution 
of $\lambda/\delta\lambda\sim2000$.  
Conditions were clear and photometric.  
The data were reduced following standard procedures (bias, flat, 
wavelength, and flux calibration) using the IRAF 
{\tt twodspec} and {\tt onedspec} packages.  
Observations of the standard star G191B2B \citep{mg90} 
were used to derive the response curve.  
Due to an observing error, we had to use a spectrum of 
G191B2B taken on a subsequent night 
through thin cirrus, so the absolute fluxes are 
inaccurate but the relative fluxes are sound.  
An approximate correction for atmospheric 
extinction was applied using the standard KPNO extinction 
curve, but we note that Kitt Peak is at a lower 
elevation (2100\,m) than Mt.\,Graham (3300\,m), so we are 
likely over-compensating at the far blue end of the spectrum.

\subsection{X-Ray Data}
We combined and analyzed all {\em Swift} X-Ray Telescope
\citep[XRT;][]{bdn05} observations obtained simultaneously
with the UVOT data between 2011-10-02 and 2012-01-17.
We searched for X-ray emission using a  4~pixel ($9\farcs6$) radius
centered on the optical position of the SN and corrected for the XRT
100\% encircled energy radius. The background was extracted locally
using a $40\arcsec$ radius source-free aperture to account for detector
and sky background, and any unresolved emission from the host.

An X-ray source is detected at the position of the SN at a $4.7\sigma$ 
statistical significance
in the merged data (88.1\,ks total exposure time).
The  PSF, deadtime, and vignetting corrected count rate of $(5.3\pm1.1) 
\times10^{-4}$\,counts\,s$^{-1}$
corresponding to an unabsorbed (0.2--10\,keV band) X-ray flux of
$f_{0.2-10} = (2.3\pm0.5) \times10^{-14}$\,erg\,cm$^{-2}$\,s$^{-1}$
and a luminosity of $L_{0.2-10} =(1.0\pm0.2) 
\times10^{39}$\,erg\,s$^{-1}$ assuming a thermal plasma
with $kT=10$~keV \citep[cf.][]{fc96,is07}, a Galactic foreground column 
density with no intrinsic
absorption of $N_H = 7.8\times10^{19}$\,cm$^{2}$ \citep{dl90}, and a 
distance of 19.2\,Mpc.
The Swift XRT X-ray image of SN 2011ht and its host galaxy is shown in 
the right-hand panel of Figure~\ref{fig-2011ht}.

Re-binning the data into two epochs with similar exposures times gives 
fluxes that are consistent within
the statistical errors of the low photon statistics. Binning the data 
into an epoch ranging from 2011-11-11
to 2012-01-17 with a total exposure time 70.4\,ks gives the highest S/N 
($5.1\sigma$) and an X-ray flux and luminosity of
$f_{0.2-10} = (3.1\pm0.6) \times10^{-14}$\,erg\,cm$^{-2}$\,s$^{-1}$ and 
a luminosity of
$L_{0.2-10} = (1.4\pm0.3) \times10^{39}$\,erg\,s$^{-1}$, respectively.

The higher flux in this late epoch and the small offset of the X-ray 
source from the optical position of the SN ($<2''$),
well within the XRT point-spread function ($15''$ half power diameter), 
confirm that the X-ray emission is from SN 2011ht,
although a contamination with unresolved X-ray sources within the host 
galaxy cannot be excluded.
Observations of previous SNe show that X-rays arise from the interaction 
of the outgoing SN shock
with substantial amounts of circumstellar material (CSM). The bulk of 
the X-ray flux is produced by the
reverse shock at low energies (around 1\,keV) as soon as the expanding 
optical shell becomes optically thin
for soft X-rays. 

\subsection{Pre-Explosion Images}
\begin{figure} 
\epsscale{0.9} 
\plotone{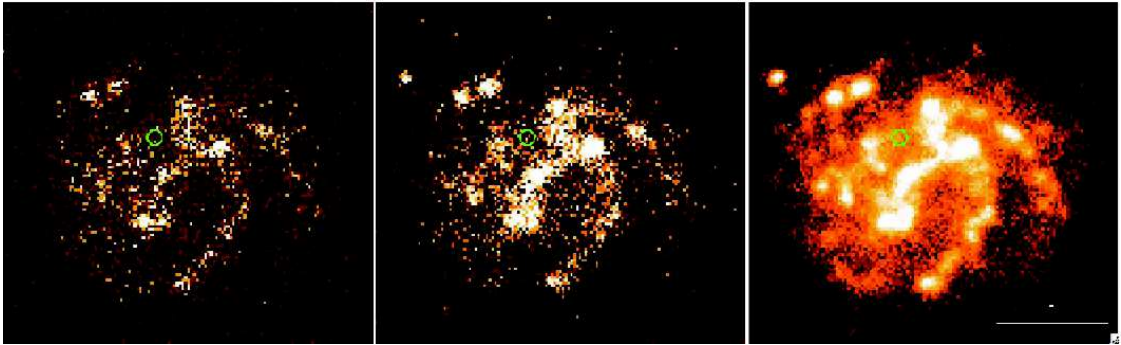} 
\caption{GALEX FUV ({\em left}) and NUV ({\em middle}) 
109\,s images of UGC~5460 taken on 2007-02-22. GALEX 
NUV ({\em right}) 1700\,s image of UGC~5460 taken on
2010-04-16. The SN position is indicated by the 
$5\arcsec$ circles.} 
\label{fig-galex}    
\end{figure}

The field of SN~2011ht was observed by the Galaxy 
Evolution Explorer \citep[GALEX;][]{mdc05} as part of 
the All Sky Imaging Survey (ASIS) in both the near-UV 
(NUV; 1800--2800\,\AA ) and far-UV (FUV; 1300--1800\,\AA) 
for a duration of 109\,s on 2007-02-22. No point source 
is detected coincident with the position of the SN 
in either band (Figure~\ref{fig-galex}). 
We measured the count rate at the SN location using a $3\farcs8$ 
(GALEX aperture No.~3) radius aperture in both the NUV 
and FUV background subtracted images and at 6 other 
locations that were close to the SN position and 
at similar isophotal brightnesses in the galaxy. Each count rate was 
converted into magnitudes using GALEX zero points 
\citep{mp07} and an aperture correction of 0.62\,mag for 
the NUV and 0.77\,mag for the FUV. We used the 
standard deviation of the fluxes in these six surrounding apertures to 
estimate the background and its uncertainties and then set 
the $3\sigma$ upper limit for detecting a 
source at the location of the SN to be the 
flux at the location of the SN minus three 
times the standard deviation. This led to $3\sigma$ 
upper limits of 19.81\,mag ($\nu L_{\nu}< 4.6\times10^6$\,L$_{\odot}$)
for the NUV and 19.66\,mag ($\nu L_{\nu}< 7.5\times10^6$\,L$_{\odot}$)
for the FUV.  The field was observed again by 
GALEX on 2010-04-16 for a duration of 1700~s
in the NUV as part of Guest Observer 
program No.~61 (Figure~\ref{fig-galex}:
{\em Right}). No point source is detected coincident 
with the position of the SN. The $3\sigma$ upper 
limit derived from this image is 19.50\,mag, consistent 
with the upper limit found from the shallower 109\,s NUV ASIS data.

\begin{figure} 
\epsscale{0.9} 
\plotone{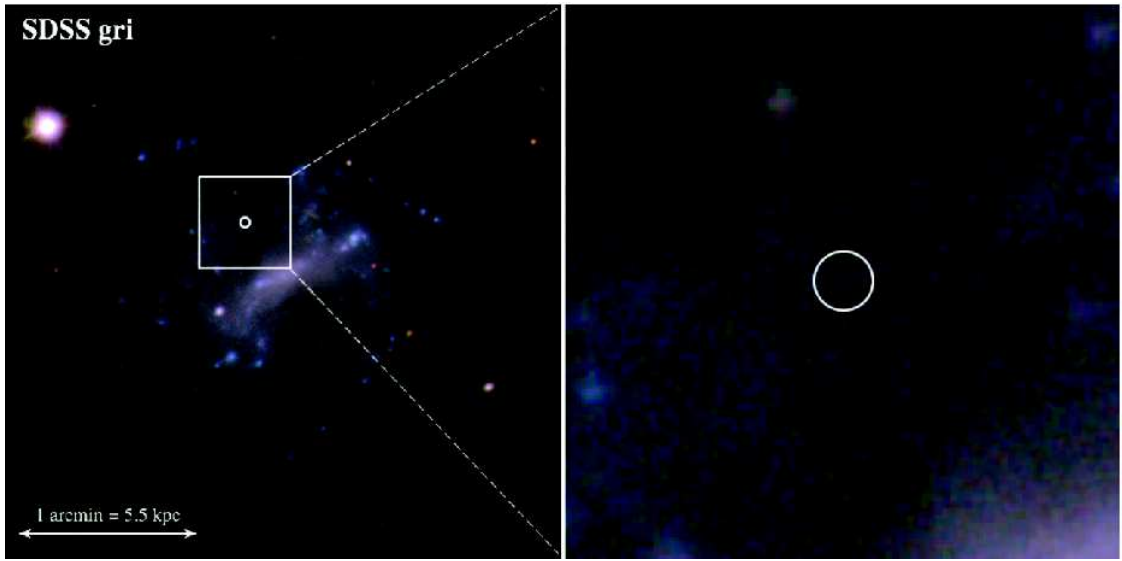} 
\caption{Pre-explosion SDSS $gri$ images of UGC~5460. 
{\em Left:} $3\arcmin \times 3\arcmin$ image centered on the 
host galaxy showing the SN location with respect to the
spiral arms. {\em Right:} $30\arcsec \times 30\arcsec$ region 
around the SN showing the environment. 
The circle at the position of the SN has a radius 
of $0\farcs6$, which 
is $5\times$\ the uncertainty in the SN position.} 
\label{fig-sdss}    
\end{figure}

We also used the Sloan Digital Sky Survey (SDSS) DR8 \citep{ah11}
pre-explosion images to search for a possible luminous progenitor  
at the position of SN~2011ht. 
First a $V$-band image of the field of SN~2011ht  
was obtained on 2011-11-01 with the Ohio State Multi-Object 
Spectrograph \citep[OSMOS;][]{mp11} mounted on the MDM 2.4-m telescope. 
The centroids of seven point 
sources around the SN position in the OSMOS 
and SDSS $g$-band images were measured, followed by 
a coordinate transformation between the two images 
using a second-order polynomial in the IRAF task {\tt geomap}. 
The resulting standard deviation is 0.3\,pixels in the 
SDSS image ($0\farcs12$). No point sources are detected in any of 
the SDSS $ugriz$ images within the error circle 
of the SN position, although there is an unresolved 
stellar background because SN~2011ht is in a crowded
environment (Figure~\ref{fig-sdss}). We analyzed 
the images with DAOPHOT \citep{spb87}
and determined a $5\sigma$ detection limit based 
on point sources found in regions with
similar background fluxes \citep[e.g.][]{lc11}. 
The resulting $5\sigma$ upper limits on the magnitude 
($\nu L_{\nu}$) are $u=21.9$ ($6.4\times10^5$\,L$_{\odot}$), 
$g=22.8$ ($2.1\times10^5$\,L$_{\odot}$), 
$r=22.4$ ($2.3\times10^5$\,L$_{\odot}$), 
$i=22.4$ ($1.8\times10^5$\,L$_{\odot}$), and 
$z=20.8$ ($6.7\times10^5$\,L$_{\odot}$). 

\begin{figure} 
\epsscale{0.9} 
\plotone{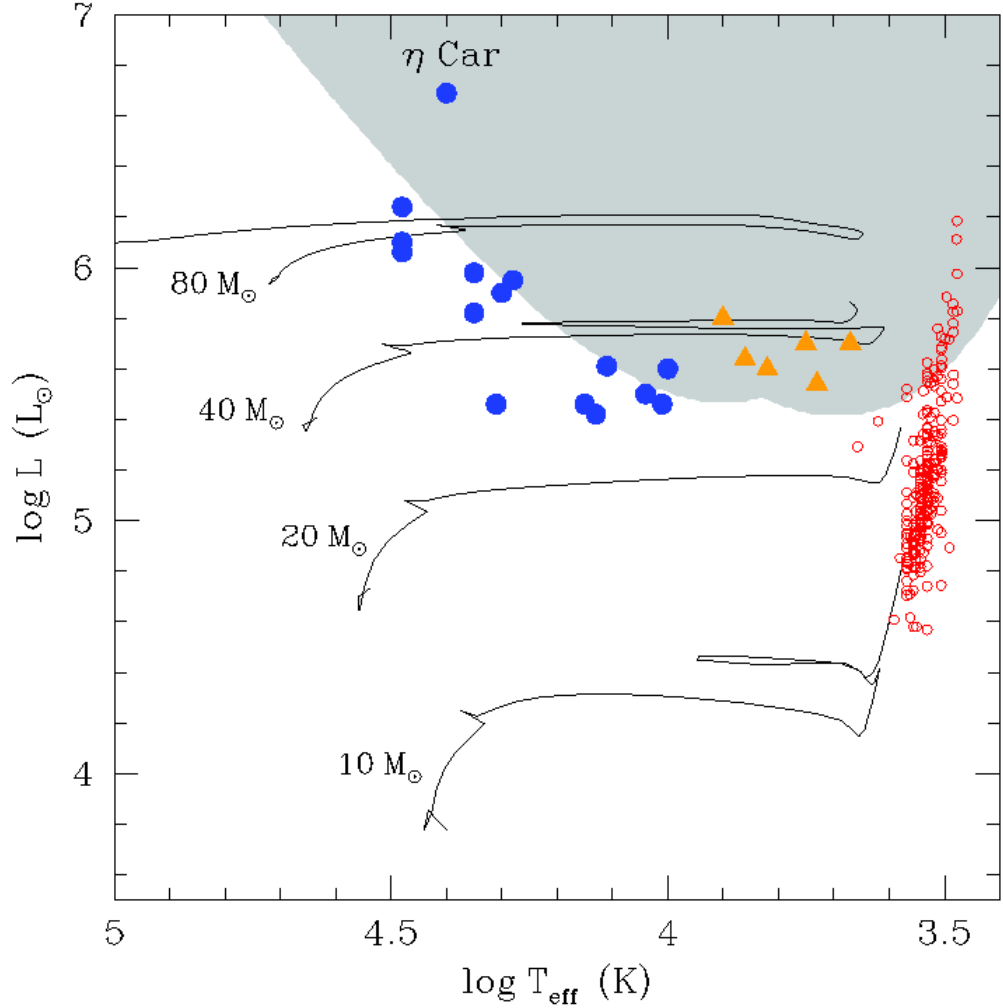} 
\caption{Constraints on the progenitor of SN~2011ht.
The grey region is ruled out by the SDSS fluxes.
The blue closed circles are known LBVs (mostly Galactic) 
in their quiescent state, orange triangles 
are yellow hypergiants, and red open circles are 
red supergiants (RSGs) in the LMC and SMC. The lines 
are evolutionary tracks for single stars 
(Z=0.004).} 
\label{fig-progen}    
\end{figure}

We compared these limits to blackbodies with a range of 
luminosities and temperatures, leading to the limits
on the progenitor shown in Figure~\ref{fig-progen}.
These limits can then be compared to the 
progenitors of known LBVs (mostly Galactic) in their 
``hot'' state \citep{hd94,djc98}, yellow hypergiants 
\citep{hd94,djc98}, red supergiants (RSGs) in the LMC and SMC 
\citep{lem06}, and the evolutionary tracks for single 
stars with masses of 10, 20, 40, and 80\,M$_{\odot}$
from the STARS models \citep{et04} for a metallicity of $Z=0.004$ 
that is consistent with the measured oxygen 
abundance of an HII region near SN~2011ht
(see below). As expected, Figure~\ref{fig-progen} reveals that
the limits are not very constraining.  
Many LBVs in their quiescent state \citep[cf.][]{hd94,sn04} 
are allowed, although something extreme like 
$\eta$-Car can be ruled out. Massive 
($\gtrsim$30\,M$_\odot$) RSGs and yellow 
hypergiants can also be excluded. 

We also retrieved the optical spectrum of UGC~5460 from SDSS DR8.  
The SDSS fiber was 
positioned on a bright \ion{H}{2} region located   
$24\arcsec$\,W and 14$\arcsec$\,N from the 
center of the galaxy  
(projected distance of $\sim$2.5\,kpc), and $36\arcsec$\,W and  
$5\arcsec$\,S 
from the location of SN~2011ht (projected distance of 
$\sim$3.4\,kpc). The spectrum is dominated by 
strong forbidden and recombination lines characteristic 
of star-forming regions, which allow us to 
estimate the metallicity of the gas. We use the $O3N2$ 
($={\rm log\{([O\,III]\,\lambda5007/H\beta)/([N\,II]\,\lambda6583/H\alpha)\}}$) 
ratio method  described in \citet{pp04} to estimate the oxygen 
abundance of the \ion{H}{2} region. This method is 
fairly insensitive to reddening because  it uses 
flux ratios of lines that are very close in wavelength. 
We measure  the total fluxes of the emission 
lines using Gaussian profiles and correct for 
the Galactic reddening. We obtain an oxygen 
abundance on the $O3N2$ scale of 
$\rm 12 + log(O/H) = (8.20\pm 0.01)$ ($\sim$1/5\,Solar), where the 
error  only reflects statistical uncertainties in the flux measurements. 
For this paper we assume that this is the 
characteristic oxygen abundance of the host 
galaxy and the SN region. The host galaxy properties 
are listed in Table~\ref{tab3}.

\section{Results}
\subsection{The Light Curve of SN~2011ht}
In the initial UVOT observations the source is detected at 
21.01\,mag in the $uvw2$ filter, 19.07\,mag in 
the $uvw1$, 17.74\,mag in the $u$-band 
\footnote{Note that the effective wavelength of the UVOT 
$u$-band filter is bluer than when used on the ground 
because it is unaffected by atmospheric absorption.},  
and $\sim$16.8\,mag in the $b$ and $v$ filters. 
Only an upper limit was found for the $uvm2$ filter.
The peak magnitudes are $\sim13.4$, $\sim13.3$, $\sim14.4$, and $\sim14.4$ 
in the UV ($uvw2$, $uvm2$, $uvw1$), $u$, $b$, and $v$ filters, 
respectively. The peak occurred $\sim$43\,days after 
discovery in the UV and $u$ bands, and after $\sim$55\,days 
in the $b$ and $v$ bands. 
The initial rise in the light curves (the 
first 20\,days) is quite sharp, particularly in the UV 
($\Delta uvw2 = 0.32$, $\Delta uvm2 = 0.32$, 
$\Delta uvw1 = 0.23$, $\Delta u = 0.18$, 
$\Delta b = 0.11$, and $\Delta v = 0.07$\,mag\,day$^{-1}$), followed by a slower 
rise to peak (over the next $\sim$20--30\,days; $\Delta uvw2 = 0.062$, 
$\Delta uvm2 = 0.064$, $\Delta uvw1 = 0.063$, 
$\Delta u = 0.045$, $\Delta b = 0.023$, 
and $\Delta v = 0.017$\,mag\,day$^{-1}$). The initial ASAS-SN observations
began $\sim$23 days after discovery. The rise in the light curve 
is 0.05\,mag\,day$^{-1}$, consistent with the average of 
the 0.07 and 0.017 UVOT $\Delta v$ rise times.
After peak, all photometric bands are observed to
fade, although the UV fades more rapidly
($\Delta uvw2 = 0.09$, $\Delta uvm2 = 0.09$, 
$\Delta uvw1 = 0.07$, $\Delta u = 0.05$, 
$\Delta b = 0.03$, and $\Delta v = 0.02$\,mag\,day$^{-1}$).
The $\Delta_{\lambda}{\rm (15)}$ for the UV and $ubv$ bands 
is $\sim$0.2 and 0.04, respectively.

\begin{figure} 
\centering
\includegraphics[width=0.75\textwidth,angle=90]{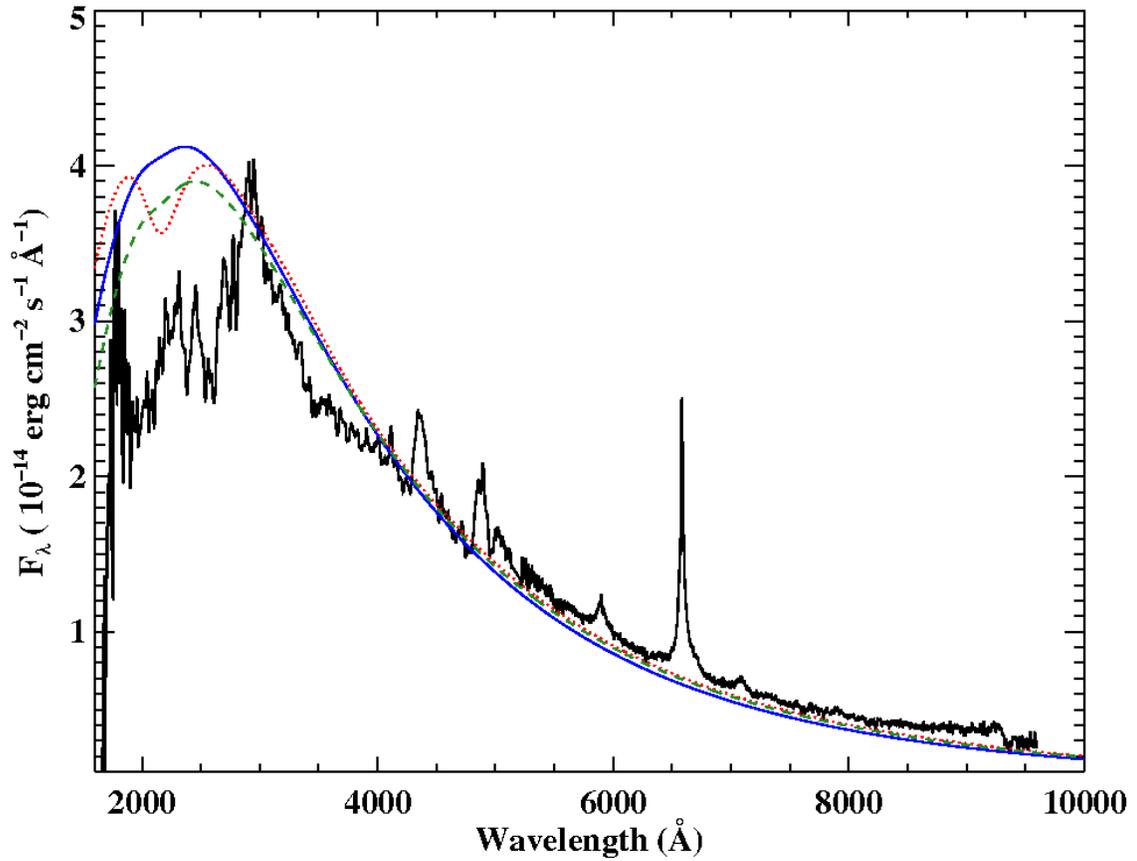} 
\caption{Blackbodies (12,900\,K) over-plotted on the combined
ARC/DIS and UVOT/uvg (Epoch 2) spectrum. The blue solid, red
dotted, and green dashed lines represent no, MW (E($B$--$V$)=0.062),
and SMC (E($B$--$V$)=0.041) host reddening, respectively.} 
\label{fig-SED}    
\end{figure}

The color evolution is peculiar because it evolves from 
red to blue at early times (Figure~\ref{fig-lc}: {\em Bottom}), 
followed by a more normal, slow reddening after the peak. This is best 
summarized by examining the evolution of the luminosity and temperature, which we can 
approximate well because of the wavelength coverage provided by the
UVOT. We fit the spectral energy distributions (SEDs) as blackbodies, 
allowing for Galactic extinction \citep[E($B$--$V$)=0.01;][]{sfd98}
as well as additional extinction in the host with a range of reddening laws 
(MW, LMC, SMC)\footnote{The UVOT $uvw2$ and $uvw1$ filters suffer from a red 
leak that can lead to complications in the analysis
of red spectra \citep{bpj10,mpa10}.  The early 
spectra of SN~2011ht are very blue, and we account for the 
red leaks in the filter response functions used for the model SEDs.}.
Our best  fit host reddening value is zero, but values up to $\sim$0.2
are consistent with the data. If we only fit the peak, where the
$S/N$ is best, we obtain  values of E($B$--$V$)=0.062 (MW) and 0.041 (SMC/LMC).  
We adopt the E($B$--$V$)=0.062 MW model for our standard reddening estimate.
In Figure~\ref{fig-SED} we compare a blackbody with 
E($B$--$V$)\,=\,0, 0.041, and 0.062 to the combined ARC/DIS 
and UVOT/uvg (Epoch 2) spectrum. The blackbody approximates the
SED reasonably well in the optical, but the fluxes overestimate
somewhat the UV due to line blanketing.

\begin{figure} 
\epsscale{0.9} 
\plotone{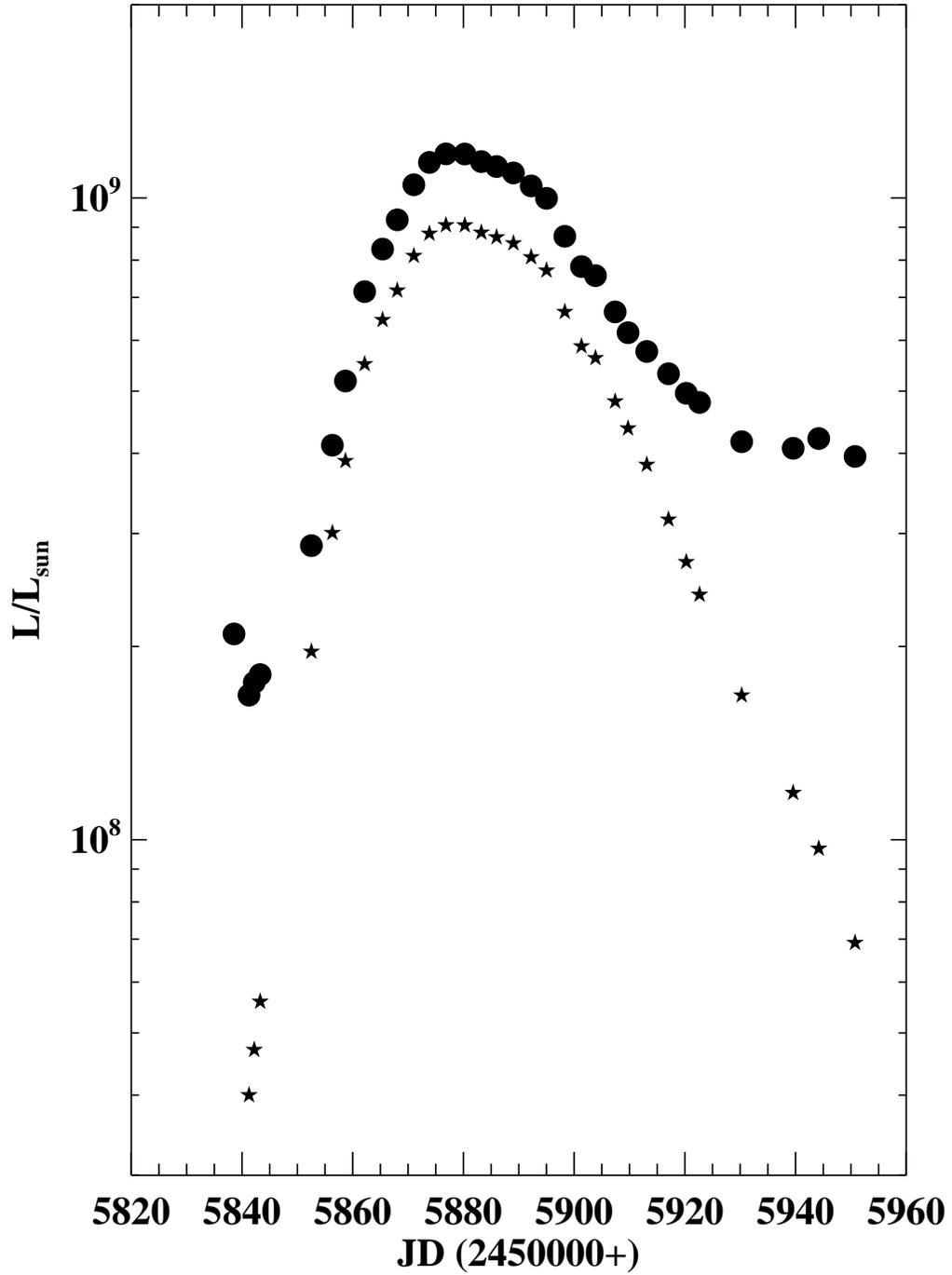} 
\caption{Bolometric (circles) and pseudo-bolometric
(stars) light curves generated from the UVOT 
photometric data. SEDs are fitted as blackbodies
to determine the bolometric luminosity. 
Pseudo-bolometric luminosities are calculated by 
summing the observed fluxes. } 
\label{fig-bol}    
\end{figure}

We also estimate a pseudo-bolometric luminosity.
We use count rate to flux conversions for the best fit 
blackbody temperatures interpolated from blackbody values in 
\citet{bpj10}, and then arrive at an average monochromatic 
flux density for each filter. The flux density is then 
integrated over the filter bandpass using trapezoidal integration.
Figure~\ref{fig-bol} shows that these two estimates of the 
luminosity agree reasonably well, with the agreement improving 
as the temperature increases and there is proportionately 
less unobserved optical/near-IR emission. The bolometric
light curve peaks at $\sim$10$^{9}$\,${\rm L_{\odot}}$, 
roughly $\sim$43\,days after discovery. The total energy emitted 
by the SN in the first 112 days is $\sim2.5 \times 10^{49}$\,ergs.
For comparison purposes, this is $\sim$100$\times$ greater-than
the energy in the eruption of the 2008~N300 transient \citep{hr11}
and $\sim$100$\times$ less-than the radiated output of
SN~2006gy \citep{sn07}.

Figure~\ref{fig-bb} shows the evolution of the blackbody fits 
in temperature ($T_{\rm{BB}}$), radius ($R_{\rm{BB}}$), and luminosity 
($L_{\rm{BB}}$) along with the fit residuals. The fits are 
worse at early times, but we see a general trend that as the 
temperature increases the luminosity increases and the black 
body radius shrinks and then remains fairly constant
until peak. After peak, $T_{\rm{BB}}$ and $L_{\rm{BB}}$ begin declining, 
while $R_{\rm{BB}}$ begins increasing. The estimates of
$T_{\rm{BB}}$, $R_{\rm{BB}}$, 
and $L_{\rm{BB}}$ for each epoch are reported in Table~\ref{tab5}.

\subsection{The Spectra of SN~2011ht}
A close examination of the UVOT/uvg spectra (Figure~\ref{fig-spect}) 
reveals absorption lines of \ion{Fe}{2} and potential 
lines of \ion{Si}{2}, \ion{Si}{3}, \ion{Ni}{2}, and \ion{N}{3}. 
The UV spectrum below 1900\,\AA\ shows 
significant absorption features which differ between the 
two spectra taken 11\,days apart. The most likely 
explanation is a combination of noise in the early spectrum 
and physical changes in the state and velocity of the 
circumstellar matter. Only the longer exposure of the later 
spectrum can be trusted below 1900\,\AA. The spectra also show broadened 
hydrogen lines of H$\beta$, H$\gamma$, and H$\delta$. 

\begin{figure} 
\epsscale{0.9} 
\plotone{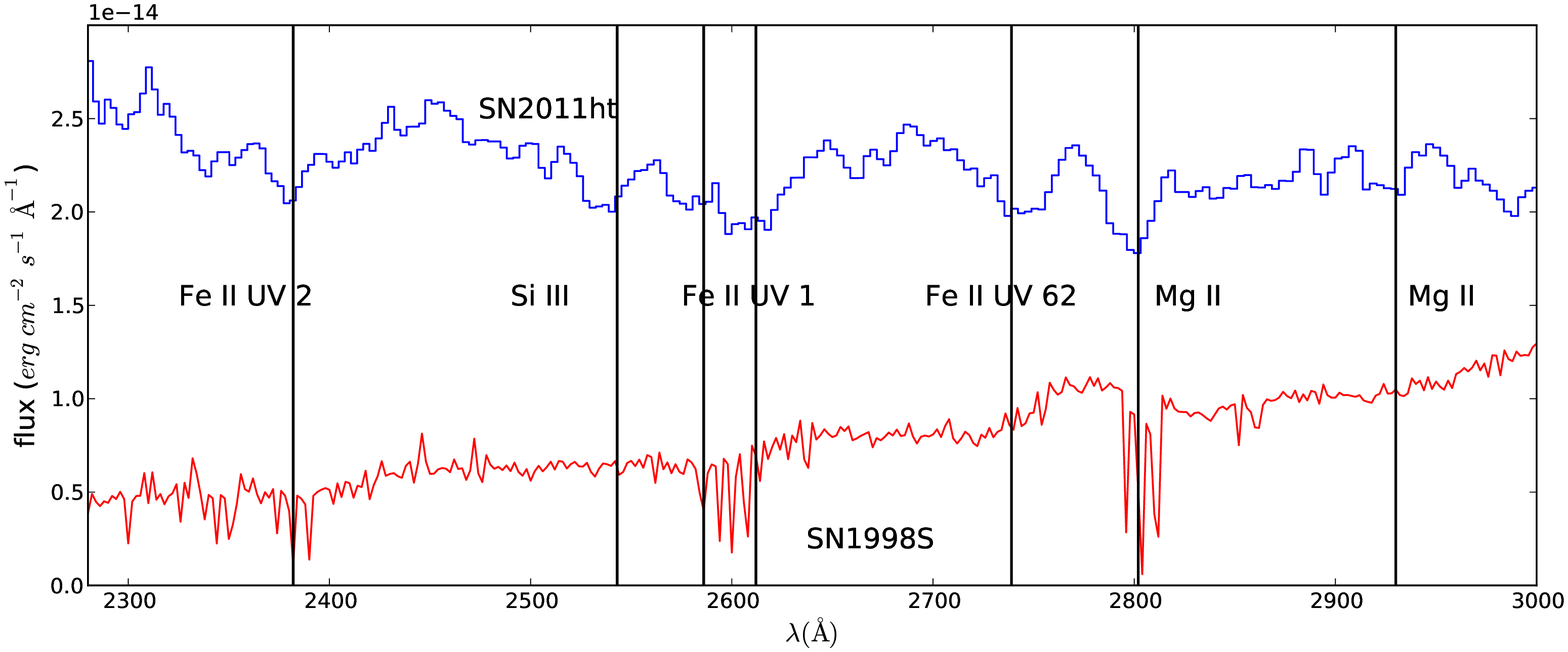}
\caption{{\em Swift} UVOT/uvg spectrum of SN~2011ht as 
compared to the spectrum of SN~1998S \citep{bdv00}.
The red spectrum is of SN~1998S and the blue is of 
SN~2011ht (UVOT/uvg Epoch 2).} 
\label{fig-gclose}    
\end{figure}

The UV is dominated by lines of \ion{Fe}{2}, most notably the UV1, UV2 and
UV62 multiplets. A comparison with the SN~1998S spectrum from \citet{bdv00} 
reveals that several sets of absorption lines match the SN~2011ht lines,
namely \ion{Mg}{2} ($\sim\lambda2800$) and \ion{Fe}{2} ($\sim\lambda2400$, 
$\lambda2586$, and $\lambda2612$; Figure~\ref{fig-gclose}). An
11\,\AA\ shift of the UVOT spectrum (which is within the wavelength accuracy 
of the instrument) was applied based on the line identifications,
so it is possible that the overall blueshift seen in the optical is masked.

At $\lambda1751$\,\AA\ there is a possible resonance line of \ion{Ni}{2}, but 
this is more likely the \ion{N}{3} ($\lambda1750$\,\AA) resonance line
that is broadened by a velocity of $\sim$1880\,km\,s$^{-1}$.
The \ion{Si}{3} $\lambda1842$, and $\lambda2562$ lines 
are not resonance lines, but are probably the
result of a cascade from \ion{Si}{4}. The line broadening of the two 
\ion{Si}{3} lines indicates velocities of 
1220 and 1474\,km\,s$^{-1}$, respectively.
We attribute the line found at $\lambda1813$\,\AA\ to
$\lambda1808$, $\lambda1816$, $\lambda1817$ triplet of \ion{Si}{2}.
The C~III] ($\lambda1909$\,\AA) line is seen at $\sim$1904\,\AA. 
The \ion{Mg}{2} resonance lines at $\lambda2797$ and $\lambda2802$\,\AA\ 
are present, while we also see the \ion{Mg}{2} ($\lambda2930$\,\AA) line 
whose lower level is the upper level of the $\lambda2802$\,\AA\ line.
Similar to the \ion{Si}{3} lines this could be due to
recombination of the \ion{Mg}{3} levels or 
due to high temperature collisional excitation.
The width of the UV lines is difficult to determine since there
are so many faint \ion{Fe}{2} lines confusing the measurements.
A FWHM between 25--33\,\AA\ is our best estimate, which equates
to a velocity of about 1670\,km\,s$^{-1}$.

\begin{figure} 
\epsscale{0.9} 
\plotone{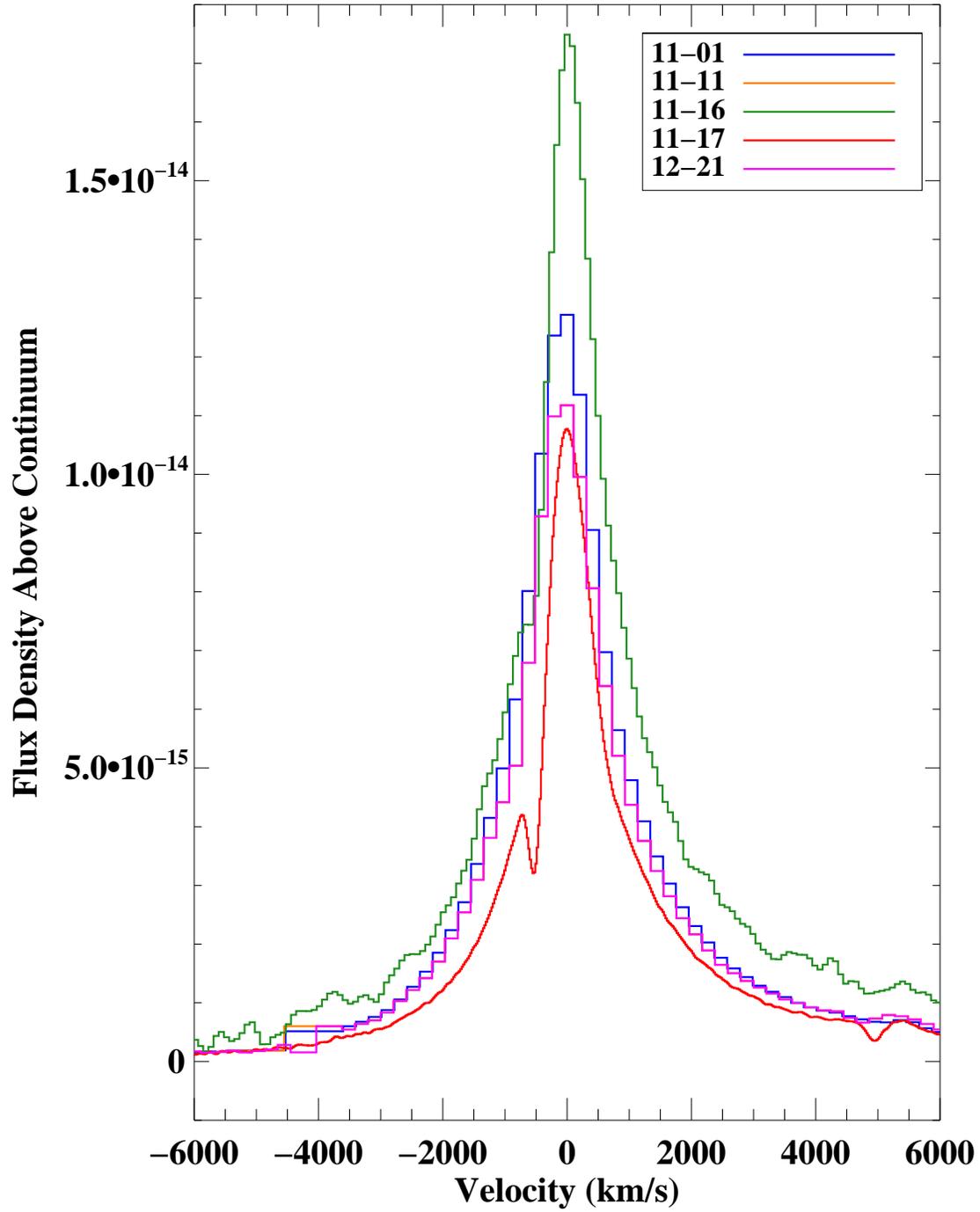} 
\caption{H$\alpha$ profile at each spectral epoch plotted
in velocity space. 
The emission peaks have been aligned to zero velocity.
The coarseness of the plots represents the resolution of
the individual instruments.} 
\label{fig-halpha}    
\end{figure}
\begin{figure} 
\epsscale{0.9} 
\plotone{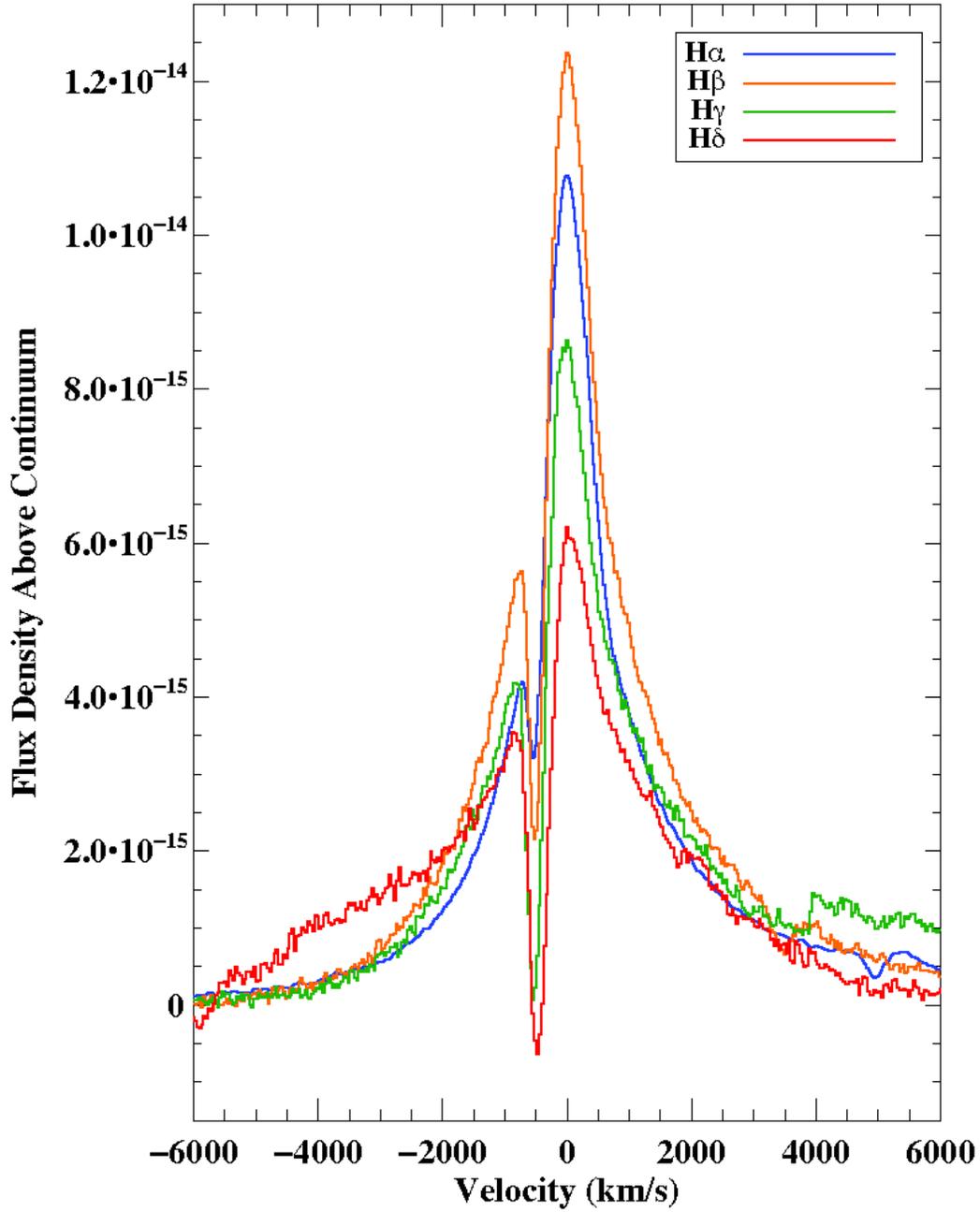} 
\caption{Balmer profiles for the LBT/MODS1 spectrum overlaid
in velocity space. The emission peaks have been aligned 
to zero velocity.} 
\label{fig-balmer}    
\end{figure}

The early optical spectra are dominated by strong 
asymmetric (the red side having greater flux) Balmer emission 
lines with narrow peaks, very broad wings, and P-Cygni profiles. 
Using the MODS1 spectrum we have measured velocities with 
respect to the various rest frame peak emission and absorption lines 
(Table~\ref{tab4}). From the H$\alpha$ P-Cygni profile, we 
measure blueshifted velocities (with respect to the velocity of
the galaxy) of 261 and 778\,km\,s$^{-1}$ in emission
and absorption, respectively. We also measure the width of the
narrow H$\alpha$ line to be 
FWHM$_{{\rm H}\alpha{\rm -Line}} \simeq 600$\,km\,s$^{-1}$ 
while the broad wings are
FWHM$_{{\rm H}\alpha{\rm -Wing}} \simeq 4200$\,km\,s$^{-1}$. 
The H$\alpha$ profile 
remains fairly consistent over the five 
epochs (see Figure~\ref{fig-halpha}). The velocity 
in absorption decreases in the higher Balmer 
lines (see Table~\ref{tab4}), and 
the absorption lines become stronger
as the emission weakens (Figure~\ref{fig-balmer}). 
H$\eta$ has no obvious emission altering the 
absorption profile, so we measure the 
absorption velocity to be 646~km\,s$^{-1}$.   
Paschen lines are easily identified on the red end 
of the spectrum. They show weak emission combined
with blue shifted absorption. 

\begin{figure} 
\epsscale{0.9} 
\plotone{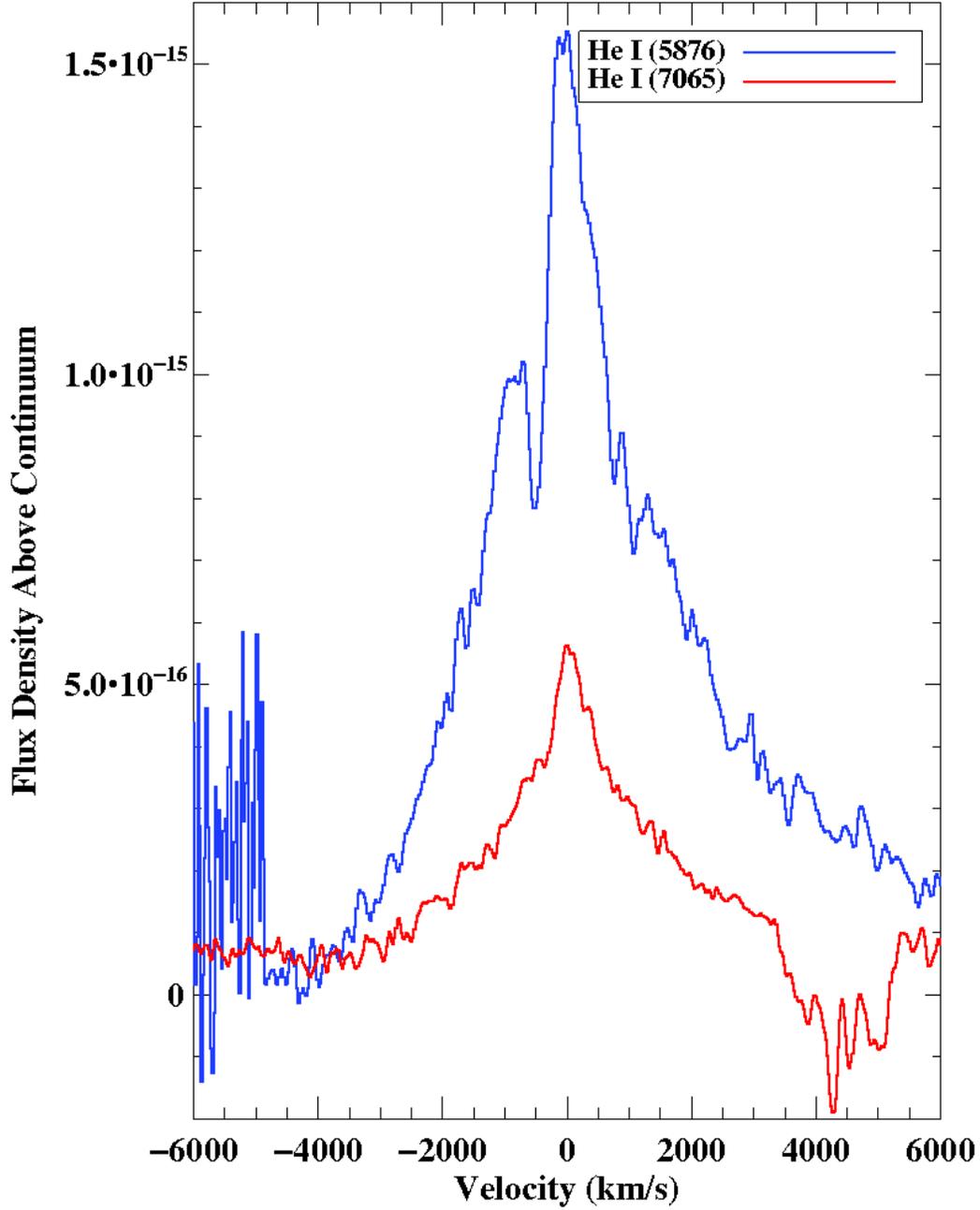} 
\caption{The \ion{He}{1} emission lines 
($\lambda7065$ and $\lambda5876$) plotted in velocity space. Note 
the broad wings in both lines
and the absorption similar to hydrogen in the 
$\lambda5876$ profile but missing in $\lambda7065$.} 
\label{fig-HeI}    
\end{figure}

There are two very strong \ion{He}{1} emission lines 
($\lambda7065$ and $\lambda5876$), both of which 
show very broad wings. The $\lambda5876$ line 
has an absorption feature similar to the hydrogen lines, but
the absorption is absent from the $\lambda7065$ 
line (Figure~\ref{fig-HeI}). The other 
\ion{He}{1} lines are weak. Normally \ion{He}{1} ($\lambda6678$) 
is a very strong line, but it lies on the red wing of 
H$\alpha$ and is difficult to analyze. The 
emission and absorption in the other 
\ion{He}{1} lines ($\lambda5015$, $\lambda4921$, 
$\lambda4471$, $\lambda4026$, and $\lambda3819$) are weaker, with  
the latter two lines having the weakest emission. 
Redward of H$\gamma$ is a peculiar continuum 
where the \ion{He}{1} ($\lambda4471$) line weakly 
appears in emission. This could be the result
of Thomson scattering.   

The \ion{O}{1} absorption triplet at $\lambda7774$ 
($\lambda7772.0$, $\lambda7774.2$, and 
$\lambda7775.4$) is strong and has a velocity of 
618~km\,s$^{-1}$, similar to the H and \ion{He}{1} absorption features. 
There are also three weak absorption lines between 3900 
and 4000\,\AA. The line measured at 3939.6\,\AA\ is most likely 
\ion{Ca}{2}~K with a velocity of $\sim 650$~km\,s$^{-1}$. 

In the last spectrum taken with the HET/LRS, the
strong, broad \ion{He}{1} emission lines
at $\lambda5876$ and $\lambda7065$ have disappeared. The other \ion{He}{1}
lines ($\lambda4921$ and $\lambda5015$) are still present 
with weak P-Cygni profiles. Several \ion{Fe}{2} lines have appeared in both 
absorption and emission. The strong absorption and P-Cygni emission is \ion{Fe}{2} 
($\lambda4924$ multiplet 42) with the next absorption 
line probably being \ion{Fe}{2} (42) at $\lambda5169$. 
There are four \ion{Fe}{2} absorption/emission lines 
from 5200--5340\,\AA\ all from multiplet 49. The 
\ion{O}{1} absorption is still present and strong.

\section{Discussion}
\begin{figure} 
\includegraphics[width=0.75\textwidth,angle=90]{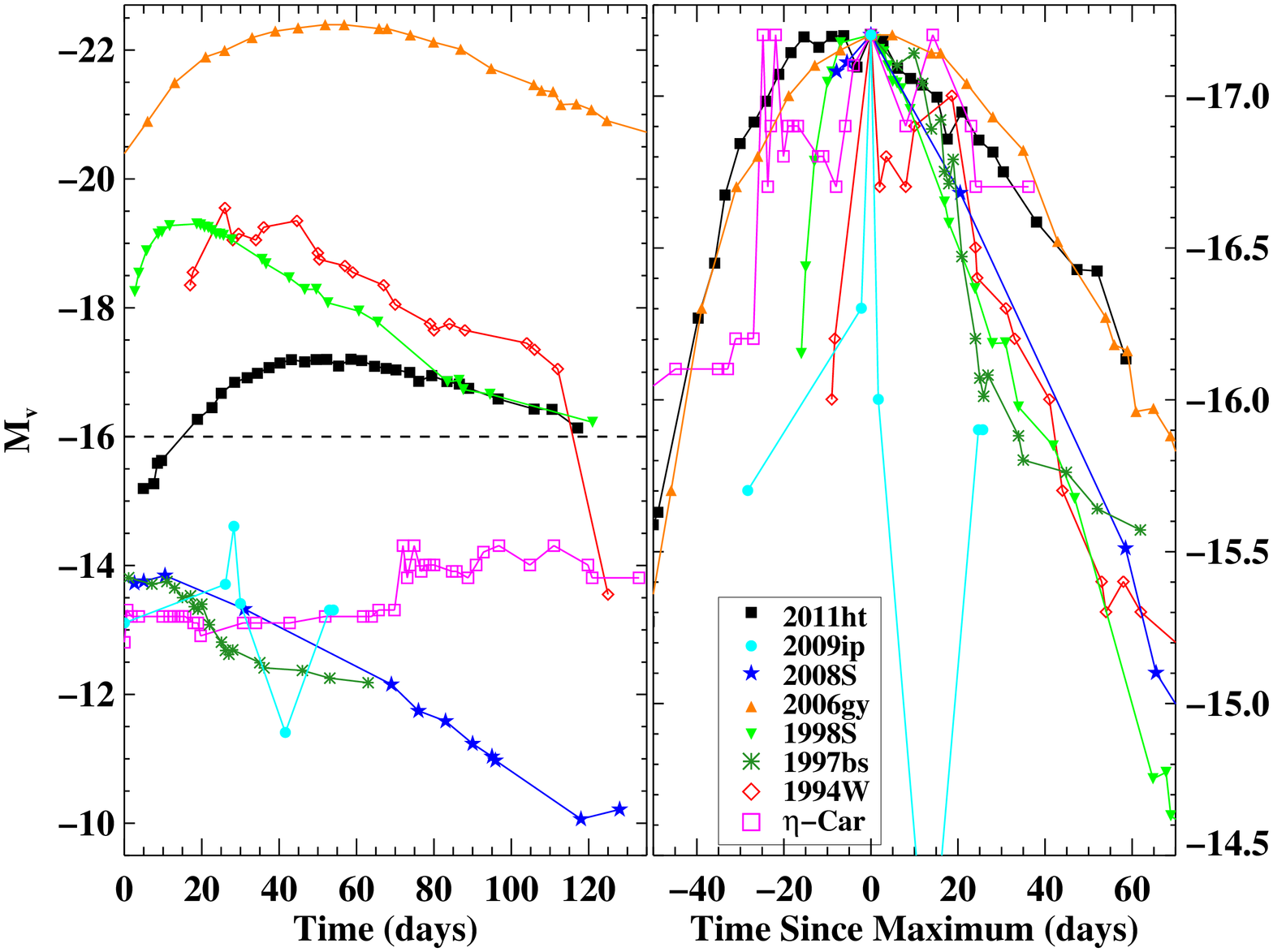} 
\caption{Comparison of the SN~2011ht light curve 
to the SNe~IIn light curves of SNe~2006gy \citep{sn07}, 
1998S \citep{lqz00}, and 1994W \citep{sj98} and the SN 
impostor light curves of SN~2009ip \citep{sn10b}, 
SN~2008S \citep{sn09}, SN~1997bs \citep{vds00} and the 1843 outburst of 
$\eta$~Car \citep{sf11}. {\em Left:} A comparison of 
the absolute magnitude assuming distance moduli for 
SNe~2011ht, 2006gy, 1998S, 1994W, 2009ip, 2008S, 1997bs, and 
$\eta$-Car of 31.42, 34.42, 31.15, 32.02, 
31.55, 28.74, 30.28, and 11.81, and an $E(B-V)$
of 0.062, 0.76, 0.15 \citep{lej01}, 0.17, 0.019, 0.64, 0.21, and 
A$_V$\,=\,1.7 \citep{dh97}, respectively. For all 
objects except $\eta$-Car the times have been shifted 
in order that the day of discovery equals zero. For 
$\eta$-Car, the shift was chosen to coincide with 
the outburst of 1843. The dotted line is the upper 
most limit for the peak magnitude of the GE-LBV sample 
by \citet{sn11}. {\em Right:} A comparison of light 
curve shapes. All light curves have been shifted so 
their peak magnitudes coincide with the peak of 
SN~2011ht. The temporal (and magnitude) shifts for 
SNe~2011ht, 2006gy, 1998S, 1994W, 2009ip, 2008S, 1997bs, and 
$\eta$-Car are 52.3 (0), 51.9 (+5.17), 18.7 
(+2.08), 26.0 (+2.33), 28.3 ($-2.61$), 10.5 ($-3.38$), 
1.1 ($-3.41$), and 72.0 ($-2.91$). The peak for $\eta$-Car was chosen 
to be the middle peak of the four major spikes. The 
light curves for SNe~2006gy and 2009ip are based 
primarily on $R$, not $V$, band data.} 
\label{fig-photocomp}    
\end{figure}

\begin{figure} 
\includegraphics[width=0.75\textwidth,angle=90]{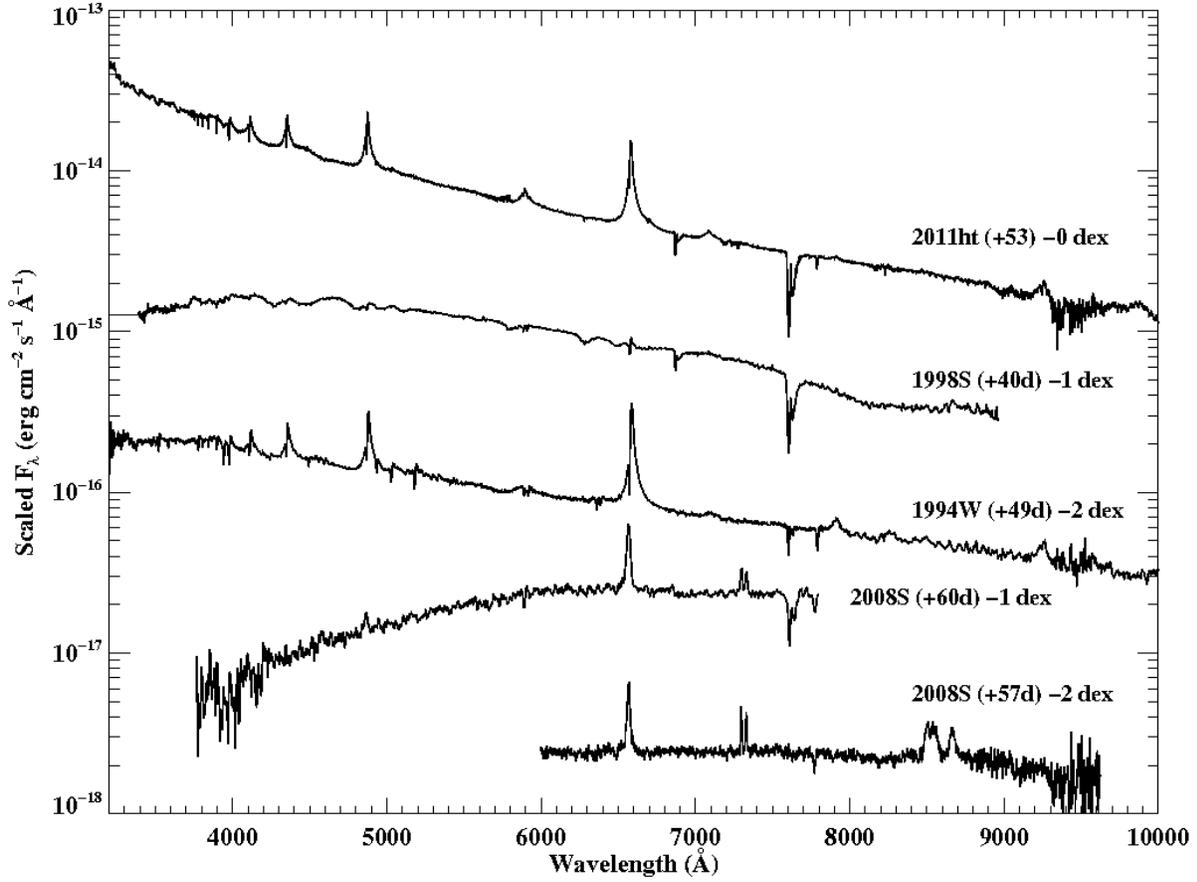} 
\caption{Comparison of the SN~2011ht spectrum to the SNe~IIn
spectra of SNe~1998S and 1994W \citep{cnn04} and the SN
impostor's spectra of SN~2008S \citep{bmt09}. The 
SN~2008S +60\,d spectrum was redshifted
by 269\AA\ to match the H$\alpha$ peak of the +57\,d spectrum.} 
\label{fig-spectracomp}    
\end{figure}

Some of the properties of SN~2011ht are similar to SN impostors
(e.g. the emission line structures), while others are comparable to 
true SNe~IIn (e.g. the luminosity) -- SN~2011ht has the luminosity
of a normal core-collapse SNe (see Figure~\ref{fig-photocomp}), 
but the spectrum of a dense wind (see Figure~\ref{fig-spectracomp}).
The spectrum originally showed hydrogen and the \ion{Ca}{2} triplet in 
emission with P-Cygni profiles, as well as narrow absorption lines of
\ion{Fe}{2} \citep{pa11}, typical of LBVs at maximum and during giant 
eruptions. The steep, blue continuum, strong
UV flux, and the properties of the \ion{He}{1} lines are
not characteristic of dense winds. The UV decay slope
of the SN~2011ht light curve is reminiscent of SN~2007pk
\citep{pta12}.

The broad asymmetric (to the red) wings in SN~2011ht are suggestive of Thomson 
scattering.  The SN impostors, SNe~1999bw and 2001ac, showed
strong, narrow Balmer lines (FWHM$_{{\rm H}\alpha-Lines} \simeq
630$ and 290\,km\,s$^{-1}$, respectively) characteristic
of LBVs, with broad wings (FWHM$_{{\rm H}\alpha-Wings} \simeq
6000$ and 3000\,km\,s$^{-1}$, respectively). The cause 
of these wings is not clear, but they have been
attributed to either electron scattering, fast 
moving ejecta, or a combination of both \citep{sn11a}.
The cause of the broad wings in SNe is not always clear
either. For SN~1998S, the broad component has been 
attributed both to the ejecta \citep{ldc00} and to 
scattering on thermal electrons \citep{cnn01}. In SN~1999W,
\citet{cnn04} attributed the broad wings to a combination 
of cool shocked gas and Thomson scattering, while \citet{dl09}
argued they were primarily the result of electron
scattering in the ejecta.  Even with detailed models it
can be difficult to reach a conclusive interpretation.
Further complicating
this picture is that some SNe~IIn initially reveal
a Thomson scattering profile but evolve
to a shocked profile \citep{sn12}.

\begin{figure} 
\epsscale{0.9} 
\plotone{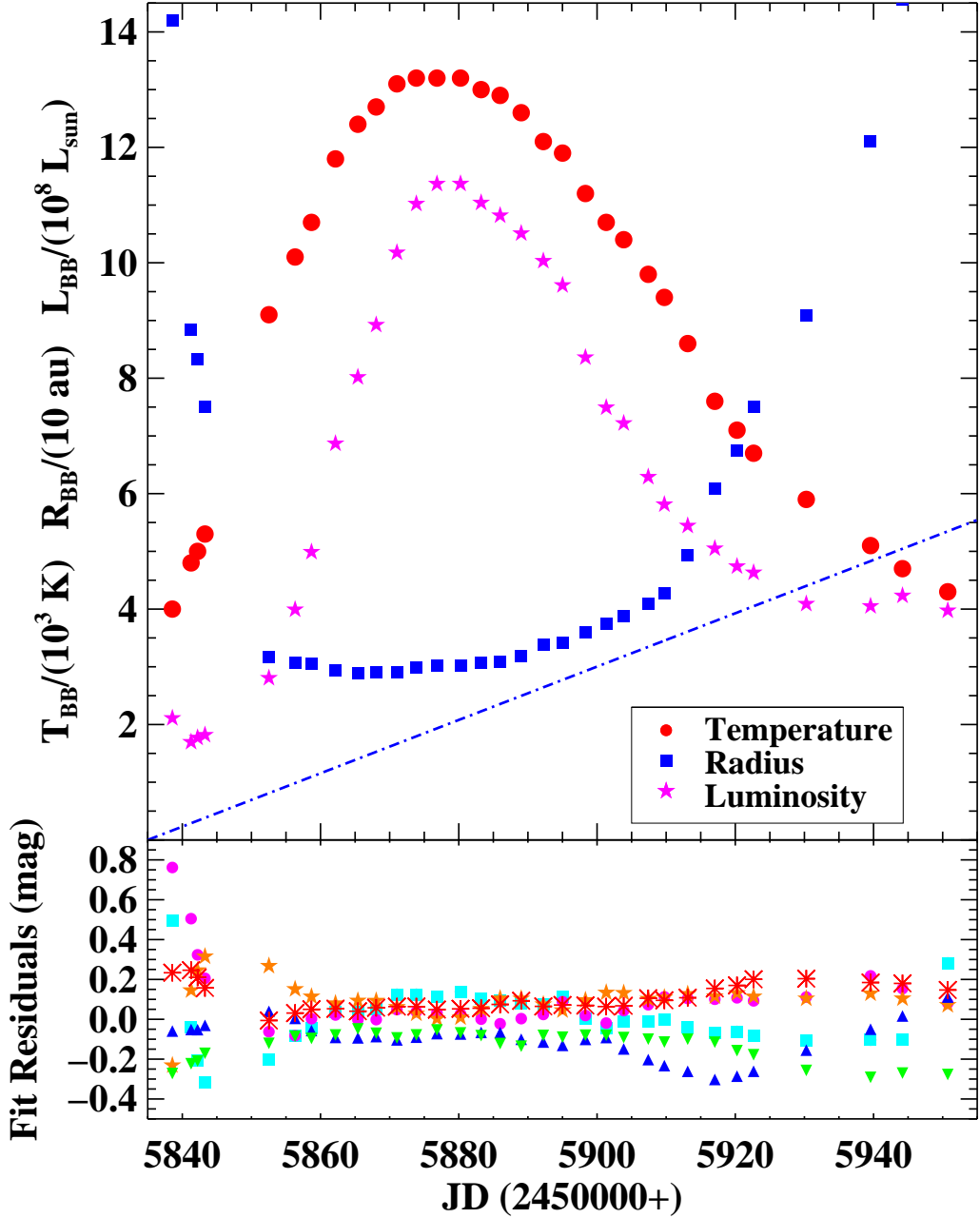} 
\caption{Blackbody temperature, radius, and luminosity with 
the corresponding residuals to the blackbody fits. The diagonal
line in the upper panel represents 800\,km\,s$^{-1}$ 
expansion.} 
\label{fig-bb}    
\end{figure}

If we suppose both the presence of Thomson scattering
in a pre-existing circumstellar medium (CSM) in order to produce
the line structures, and that the transient is a true
SN based on its energetics, then we would expect
a shock wave to be propagating through the CSM with
observable effects.  Significantly modifying the 
line structures requires $\tau_T \simeq 1$ on the
scale of the apparent photosphere (Figure~\ref{fig-bb}),
and we use this to normalize the model by
$$
  \dot{M} = { 4 \pi v_w R_{in} \tau_T \over \kappa_T }
     \simeq 0.02 \tau_T \left( { v_w \over 600~\hbox{km/s} }\right)
                 \left({ R_{in} \over 5 \times 10^{14}~\hbox{cm} }\right) 
              \left( { \Delta R \over R_{out}} \right) M_\odot~\hbox{year}^{-1}.
$$
where the Thomson opacity is $\kappa_T \simeq 0.34$ and we 
identify the absorption feature of the spectra with the wind
speed ($v_w$).  The shell extends from $R_{in}$ to $R_{out}$ with
a thickness $\Delta R = R_{out}-R_{in}$.   The term 
$\Delta R/R_{out} \rightarrow 1$ for a thick shell.
Producing significant levels of Thomson scattering
on these large scales requires a very dense wind, similar
to that of an LBV in eruption \citep{kcs11b}.
We can then estimate the H$\alpha$ luminosity assuming it is due to
recombination outside radius $R_{in}$ (and ignoring $R_{out}$) as
\begin{equation}
   L_{H\alpha} \simeq { 4 \pi E_{H\beta} \alpha_{H\beta} R_{in} \tau_T^2 \over
           \mu^2 m_p^2 \kappa_T^2 }
       \simeq 4 \times 10^6 \left( { R_{in} \over 5 \times 10^{14}~\hbox{cm}}\right)
          \tau_T^2 L_\odot 
\end{equation}
where $E_{H\alpha}=1.9$~eV, $\alpha_{H\alpha}\simeq 10^{-13}$~cm$^3$/s
is the H$\alpha$ rate, and $\mu \simeq 0.6$ is the mean molecular weight.
For the estimated H$\alpha$ line flux of $5 \times 10^{-13}$~ergs~cm$^{-2}$~s$^{-1}$
($6 \times 10^6 L_\odot$) in the MODS spectrum,  the line formation radius
will be consistent with the photometric radius for $\tau_T \simeq 1$,
roughly as needed to broaden the lines with Thomson scattering.  

A shock propagating through such a medium releases
a significant amount of energy, with a shock luminosity of
$$
   L_s \simeq { \dot{M} v_s^3 \over 2 v_w } =
         { 2 \pi \tau_T R_{in} v_s^3 \over \kappa_T }
          \left( { \Delta R \over R_{out} }\right)
       \simeq 2 \times 10^7 \tau_T \left( { v_s \over 2000~\hbox{km/s} }\right)^3
                 \left({ R_{in} \over 4\times 10^{14}~\hbox{cm} }\right)
              \left( { \Delta R \over R_{out}} \right)~L_\odot
$$
\citep[e.g.][]{maa10}, where we have set the shock velocity to 
2000\,km\,s$^{-1}$, an intermediate velocity between the narrow
and broad components.  While this energy can
be lost to re-accelerating the expansion of the flow, on
these physical scales it will tend to be radiated.  The
natural shock temperature is $k T_s \simeq 5$~keV. If
the optical depth of the wind is low, the X-rays from the
shock cannot be thermalized and simply escape.  Averaging 
over the X-ray epochs from 11 November to 17 January
(days 43-110) gives a detection with $L_X = (3.9\pm0.8) \times 10^5 L_\odot$.
These can be reconciled by having a low optical depth, $\tau_T \simeq 0.02$ 
corresponding to a mass loss rate of a few $10^{-4}M_\odot$~year$^{-1}$.  
A possible difficulty is that at the low shock densities implied 
by this optical depth, one would be directly observing the shock and
would expect to see the characteristic emission lines associated with
fast shocks \citep[e.g. the late time spectra of SN~1980K;][]{Fesen1994},
although they would be heavily diluted by the large ratio between
the shock and photospheric luminosities.

\begin{figure} 
\epsscale{0.9} 
\plotone{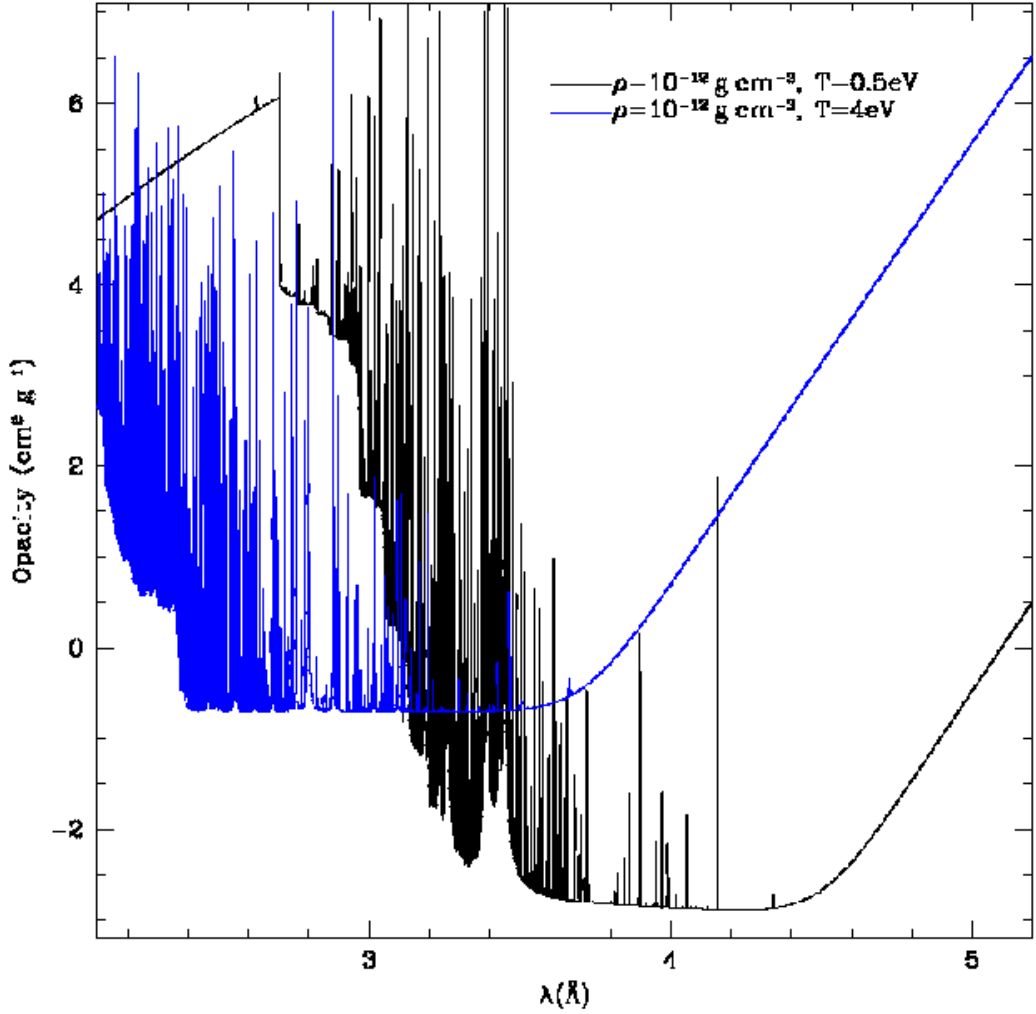} 
\caption{Wavelength dependent opacity for a 
density of 10$^{-12}$\,g\,cm$^{-3}$ 
at two different temperatures: 4\,eV (blue), 0.5\,eV (black).  As 
the shell drops below a temperature of 1\,eV, recombination 
drastically increases the opacity below a few thousand \AA 
while decreasing the electron scattering opacity (as the 
number of free electrons diminishes).  In this way, 
the X-ray optical depth can be very high while maintaining 
a modest optical depth in the optical bands.  The opacities 
are from the LANL OPLIB database \citep{mnh95}.}
\label{fig-opac}    
\end{figure}

The alternative is to make the medium dense enough so that the X-rays
are absorbed before they escape \citep[see the recent discussion
in][]{Chevalier2012}.  At the densities and temperatures we expect to
see in the shell, the material is likely to be incompletely ionized,
which means that the optical opacity (dominated by electron scattering)
will be low compared to the X-ray opacity (e.g. Figure~\ref{fig-opac}).  
In these conditions, $\tau_{\rm X-rays} \gg
\tau_{\rm T}$ and, if we require $\tau_{\rm T} \approx 1$ to fit the line
profiles, the high X-ray optical depth will be sufficient to severely
limit the softer X-rays to which {\it Swift} is most sensitive.  There
is an upper bound of $\tau_T \ltorder 100$ where $L_s$ begins to
exceed the bolometric luminosity of the transient.  If we follow the
argument of \citet{Chevalier2012} and try to produce all the
luminosity from an embedded shock, the energy available to heat the
medium is $L_s \Delta R\tau/c$, where $t_d \simeq 3\Delta R \tau/c$ is
how long it takes the energy to escape.  We equate this with the
radiative energy $4\pi \Delta R R_{in}^2 a T^4$ to estimate a
temperature
\begin{equation}
     T = \left( { 3 L_s \tau \over 16 \pi R_{in}^2 \sigma }\right)^{1/4} 
      \simeq 11000 \left( { L_s \over 10^9 L_\odot } \right)^{1/4}
                  \left( { 5 \times 10^{14}~\hbox{cm} \over R_{in} } \right)^{1/2}
                  \tau^{1/4}~\hbox{K}.
\end{equation}
Thermalizing the shock photons to limit the X-ray transmittance places
constraints on the ionization state of the atoms.  To explain the line
profiles, we require $\tau_T \sim 1$.  But the optical depth
cannot be too high, or the photon diffusion time becomes too long and
the shock breaks out before the radiation.  This also implies that
$\tau_T \ltorder 100$.  The mass associated with the shell is then
\begin{equation}
        M =  4 \pi {\tau_T \over \kappa_T } R_{in} R_{out}
       \simeq 0.01 \tau_T 
          \left( { R_{in}  \over 5 \times 10^{14}~\hbox{cm} }\right)
          \left( { R_{out} \over 10^{15}~\hbox{cm} }\right)~M_\odot
\end{equation}
where we have used $\kappa_T=0.34$~cm$^2$~g$^{-1}$ for a largely ionized 
medium, but in can be lower if the H/He are able to partially recombine
(see Figure~\ref{fig-opac}).  Theses masses and mass loss
rates in these high luminosity models are typical of other
Type~IIn SNe \citep[e.g.][]{km10} and are in the range
of LBVs in eruption \citep[e.g.][]{kcs11}.

There is one additional attractive feature of surrounding an explosive
transient with a discontinuous shell of material.  Namely, it 
might be able to explain the large photospheric
radius and low temperature of the first few epochs (see Figure~\ref{fig-bb}).
When the shock breaks out of the surface of the star, it produces
a luminosity UV/X-ray spike in excess of $10^{40}\,{\rm ergs \, s^{-1}}$ in 
a brief burst.  The duration depends on the structure of 
the star, wind profile and the photon energy, but the bulk of the 
breakout emission occurs well within 10,000\,s \citep[e.g.][]{fryer09,frey12}.
In a shell geometry, this short radiation spike rapidly crosses to the
shell and then photoionizes it before escaping.  But if the radiation
interacts in the shell, it is spread over the diffusion time $ t_d
\simeq 3 \tau \Delta R/c \simeq \tau(\Delta
R/10^{15}~\hbox{cm})$~days, which is relatively well-matched to the
duration and apparent photospheric scales of the initial phase.  
The problem is that the required
luminosity is at the high end of what we might expect from shock
breakout.  This can be explored further once the nature of this 
transient is fully understood.

\section{Conclusions}
SN~2011ht shows a series of curious behaviors.  It has a spectrum broadly
resembling a dense stellar wind but the overall luminosity and energetics
of a true core-collapse SN.  It appeared to have a very large, cool
photosphere initially, which then shrinks to roughly a constant
before starting to slowly expand.  The apparent temperature rises
dramatically from $\sim 5000$~K to almost $14,000$~K before starting 
to decline.   At early times there seems to be no X-ray emission, and
then it is detected only to probably fade again.

Clearly continued photometric and spectroscopic monitoring of this system
is necessary.  Combining a shock from an explosive transient with a dense
stellar wind has the prospect of explaining some of the peculiarities, but
such shocks have to eventually emerge at low optical depths which should
dramatically change the spectroscopic and X-ray properties as the transient
evolves.

\acknowledgments
JLP acknowledges support from NASA through 
Hubble Fellowship Grant HF-51261.01-A awarded 
by STScI, which is operated by AURA, Inc. 
for NASA, under contract NAS 5-2655.
CSK, BJS, DMS and KZS are supported 
by NSF grant AST-0908816.  KZS and RS
are also supported by NSF grant AST-1108687.
JFB was supported by NSF grant PHY-1101216.
We gratefully acknowledge the contributions from 
members of the {\em Swift} UVOT team at the Pennsylvania 
State University (PSU), University College 
London/Mullard Space Science Laboratory (MSSL), 
and NASA/Goddard Space Flight Center. This work 
is sponsored at PSU by NASA contract NAS5-00136 
and at MSSL by funding from the Science and 
Technology Facilities Council (STFC) and the UK Space Agency.
The ASAS-SN commissioning observations were only possible due 
to help and  support of the Las Cumbres Observatory, especially 
W. Rosing, E. Hawkins,  R. Ross, M. Elphick, D. Mullins and Z. Walker.
We thank R. McMillan and G. Bakos for obtaining a spectrum with 
the  APO 3.5-m telescope, and the APO director S. 
Hawley for granting DD time for this observation.
This paper used data taken with the LBT/MODS1 spectrographs built 
with funding from NSF grant AST-9987045 and the NSF Telescope 
System Instrumentation Program (TSIP), with additional funds 
from the Ohio Board of Regents and the Ohio State University Office of Research. 
The Hobby Eberly Telescope (HET) is a joint project of 
the University of Texas at Austin, PSU, 
Stanford University, Ludwig-Maximilians-Universit\"at 
M\"unchen, and Georg-August-Universit\"at G\"ottingen. The HET 
is named in honor of its principal benefactors, William P. 
Hobby and Robert E. Eberly. The Marcario Low Resolution 
Spectrograph (LRS) is named for Mike Marcario of High Lonesome Optics
who fabricated several optics for the instrument but died 
before its completion. The LRS is a joint project of the 
HET partnership and the Instituto de 
Astronomia de la Universidad Nacional Aut\'onoma de M\'exico.
Based in part on observations made with the Large Binocular Telescope.
The LBT is an international collaboration among institutions in the
United States, Italy and Germany. The LBT Corporation partners are:
the University of Arizona on behalf of the Arizona university system;
the Istituto Nazionale di Astrofisica, Italy; the LBT
Beteiligungsgesellschaft, Germany, representing the Max Planck
Society, the Astrophysical Institute Potsdam, and Heidelberg
University; the Ohio State University; and the Research Corporation,
on behalf of the University of Notre Dame, University of Minnesota and
This research has made use of the NASA/IPAC Extragalactic 
Database (NED) which is operated by the Jet Propulsion 
Laboratory, California Institute of Technology, under 
contract with the National Aeronautics and Space 
Administration.

{\it Facilities:} \facility{Swift (UVOT)}, \facility{ARC (DIS)}, 
\facility{HET (LRS)}, \facility{LBT (MODS1)}

\clearpage

\clearpage

\clearpage

\clearpage

\clearpage

\clearpage

\clearpage

\clearpage

\clearpage

\clearpage

\clearpage

\clearpage



\begin{deluxetable}{cccccccc} 
\tablecolumns{8} 
\tabletypesize{\scriptsize} 
\tablecaption{{\em Swift} UVOT and ASAS-SN photometry of SN~2011ht  
\label{tab1}} 
\tablewidth{0pt} 
\tablehead{ 
  \colhead{UT} &
  \colhead{Instrument} & 
  \multicolumn{6}{c}{Observed Magnitudes$^{{\rm a}}$}\\ 
  \colhead{(JD2450000+)} &
  \colhead{} & 
  \colhead{$uvw2$} & 
  \colhead{$uvm2$} & 
  \colhead{$uvw1$} & 
  \colhead{$u$} & 
  \colhead{$b$} & 
  \colhead{$v$/$V^{{\rm b}}$}  
}
\startdata
5838.6  & UVOT    & 21.01(46) & 20.24(**) & 19.07(16) & 17.74(09) & 17.14(07) & 16.45(07)\\
5841.3  & UVOT    & 20.37(42) & 20.39(**) & 18.42(14) & 16.95(09) & 16.72(07) & 16.37(09)\\
5842.2  & UVOT    & 19.37(21) & 19.80(27) & 18.32(14) & 16.69(09) & 16.62(07) & 16.05(08)\\
5843.3  & UVOT    & 18.68(14) & 19.13(18) & 18.01(12) & 16.45(08) & 15.41(07) & 16.01(08)\\
5852.5  & UVOT    & 15.70(15) & 15.50(09) & 15.40(10) & 14.84(11) & 15.50(10) & 15.37(09)\\
5856.3  & UVOT    & 15.13(17) & 14.94(15) & 14.78(12) & 14.43(11) & 15.25(09) & 15.19(07)\\
5857.07 & ASAS-SN &    --     &    --     &    --     &    --     &    --     & 15.00(02)\\
5857.08 & ASAS-SN &    --     &    --     &    --     &    --     &    --     & 15.04(02)\\
5857.09 & ASAS-SN &    --     &    --     &    --     &    --     &    --     & 15.01(02)\\
5857.10 & ASAS-SN &    --     &    --     &    --     &    --     &    --     & 15.03(02)\\
5858.08 & ASAS-SN &    --     &    --     &    --     &    --     &    --     & 14.97(02)\\
5858.09 & ASAS-SN &    --     &    --     &    --     &    --     &    --     & 14.96(02)\\
5858.10 & ASAS-SN &    --     &    --     &    --     &    --     &    --     & 14.93(02)\\
5858.11 & ASAS-SN &    --     &    --     &    --     &    --     &    --     & 14.94(02)\\
5858.12 & ASAS-SN &    --     &    --     &    --     &    --     &    --     & 14.94(02)\\
5858.13 & ASAS-SN &    --     &    --     &    --     &    --     &    --     & 14.93(02)\\
5858.14 & ASAS-SN &    --     &    --     &    --     &    --     &    --     & 14.96(02)\\
5858.7  & UVOT    & 14.67(20) & 14.46(11) & 14.35(14) & 14.09(12) & 15.02(10) & 14.97(07)\\
5859.06 & ASAS-SN &    --     &    --     &    --     &    --     &    --     & 14.96(02)\\
5859.07 & ASAS-SN &    --     &    --     &    --     &    --     &    --     & 14.96(02)\\
5859.09 & ASAS-SN &    --     &    --     &    --     &    --     &    --     & 14.92(02)\\
5859.10 & ASAS-SN &    --     &    --     &    --     &    --     &    --     & 14.94(02)\\
5859.11 & ASAS-SN &    --     &    --     &    --     &    --     &    --     & 14.98(02)\\
5862.06 & ASAS-SN &    --     &    --     &    --     &    --     &    --     & 14.76(01)\\
5862.07 & ASAS-SN &    --     &    --     &    --     &    --     &    --     & 14.74(01)\\
5862.08 & ASAS-SN &    --     &    --     &    --     &    --     &    --     & 14.74(01)\\
5862.10 & ASAS-SN &    --     &    --     &    --     &    --     &    --     & 14.76(01)\\
5862.11 & ASAS-SN &    --     &    --     &    --     &    --     &    --     & 14.74(01)\\
5862.13 & ASAS-SN &    --     &    --     &    --     &    --     &    --     & 14.70(01)\\
5862.14 & ASAS-SN &    --     &    --     &    --     &    --     &    --     & 14.73(01)\\
5862.15 & ASAS-SN &    --     &    --     &    --     &    --     &    --     & 14.73(01)\\
5862.2  & UVOT    & 14.26(23) & 14.05(19) & 13.97(16) & 13.74(12) & 14.77(10) & 14.80(08)\\
5864.06 & ASAS-SN &    --     &    --     &    --     &    --     &    --     & 14.69(01)\\
5864.07 & ASAS-SN &    --     &    --     &    --     &    --     &    --     & 14.67(01)\\
5864.08 & ASAS-SN &    --     &    --     &    --     &    --     &    --     & 14.70(01)\\
5864.11 & ASAS-SN &    --     &    --     &    --     &    --     &    --     & 14.64(01)\\
5864.12 & ASAS-SN &    --     &    --     &    --     &    --     &    --     & 14.64(01)\\
5864.13 & ASAS-SN &    --     &    --     &    --     &    --     &    --     & 14.66(01)\\
5864.14 & ASAS-SN &    --     &    --     &    --     &    --     &    --     & 14.64(01)\\
5864.15 & ASAS-SN &    --     &    --     &    --     &    --     &    --     & 14.64(02)\\
5865.4  & UVOT    & 13.96(27) & 13.73(21) & 13.62(19) & 13.55(13) & 14.66(11) & 14.73(08)\\
5868.1  & UVOT    & 13.89(21) & 13.62(19) & 13.67(14) & 13.52(10) & 14.64(09) & 14.66(07)\\
5871.1  & UVOT    & 13.67(30) & 13.45(28) & 13.38(20) & 13.36(14) & 14.44(11) & 14.57(08)\\
5873.9  & UVOT    & 13.68(24) & 13.41(23) & 13.29(17) & 13.27(12) & 14.41(10) & 14.50(08)\\
5876.8  & UVOT    & 13.55(27) & 13.31(25) & 13.21(19) & 13.27(12) & 14.37(10) & 14.45(08)\\
5880.3  & UVOT    & 13.62(30) & 13.37(28) & 13.23(21) & 13.28(13) & 14.39(11) & 14.48(08)\\
5883.2  & UVOT    & 13.64(30) & 13.35(26) & 13.28(21) & 13.29(13) & 14.36(11) & 14.45(08)\\
5886.0  & UVOT    & 13.66(29) & 13.38(27) & 13.40(19) & 13.32(12) & 14.37(11) & 14.44(08)\\
5889.0  & UVOT    & 13.71(22) & 13.47(20) & 13.44(15) & 13.31(11) & 14.32(09) & 14.55(07)\\
5892.2  & UVOT    & 13.81(28) & 13.55(26) & 13.47(19) & 13.30(13) & 14.39(11) & 14.44(08)\\
5895.0  & UVOT    & 13.96(27) & 13.71(23) & 13.55(19) & 13.36(13) & 14.44(11) & 14.46(08)\\
5898.3  & UVOT    & 14.09(19) & 13.87(17) & 13.66(14) & 13.43(10) & 14.45(09) & 14.55(07)\\
5901.4  & UVOT    & 14.76(12) & 14.27(11) & 14.06(12) & 13.67(10) & 14.63(08) & 14.59(07)\\
5903.9  & UVOT    & 14.48(23) & 14.24(21) & 14.02(17) & 13.62(14) & 14.60(11) & 14.61(08)\\
5907.4  & UVOT    & 14.70(16) & 14.53(13) & 14.22(12) & 13.64(10) & 14.66(09) & 14.65(07)\\
5909.8  & UVOT    & 15.02(18) & 14.81(16) & 14.45(14) & 13.84(12) & 14.80(10) & 14.78(08)\\
5913.1  & UVOT    & 15.32(13) & 15.14(11) & 14.68(10) & 13.88(10) & 14.78(09) & 14.70(07)\\
5917.0  & UVOT    & 15.83(11) & 15.64(09) & 15.09(09) & 14.08(09) & 14.92(08) & 14.79(07)\\
5920.2  & UVOT    & 16.38(10) & 16.20(08) & 15.45(08) & 14.30(09) & 14.96(08) & 14.83(07)\\
5922.6  & UVOT    & 16.53(10) & 16.37(08) & 15.63(08) & 14.45(09) & 15.03(08) & 14.89(07)\\
5930.3  & UVOT    & 17.43(10) & 17.44(08) & 16.32(08) & 15.05(08) & 15.25(08) & 15.06(07)\\
5939.6  & UVOT    & 18.36(12) & 18.54(12) & 17.15(09) & 15.77(08) & 15.58(07) & 15.21(07)\\
5944.2  & UVOT    & 18.77(15) & 19.26(19) & 17.53(10) & 16.07(08) & 15.73(07) & 15.22(07)\\
5950.7  & UVOT    & 19.96(29) & 19.95(**) & 18.18(12) & 16.79(08) & 16.15(07) & 15.50(07)\\
\enddata 
\tablenotetext{a}{The first two $uvm2$
values are $2\sigma$-level detections.
Values in parenthesis are the errors. 
$3\sigma$ upper limits are marked with **.} 
\tablenotetext{b}{The $v$/$V$ column indicates
the UVOT/ASAS-SN filter.}
\end{deluxetable} 

\begin{deluxetable}{lcr} 
\tablecolumns{3} 
\tabletypesize{\scriptsize} 
\tablecaption{Spectroscopic Observations of SN~2011ht 
\label{tab2}} 
\tablewidth{0pt} 
\tablehead{ 
  \colhead{Instrument} & 
  \colhead{Date} & 
  \colhead{Exposure}\\ 
  \colhead{} & 
  \colhead{(UT)} & 
  \colhead{(s)} 
}
\startdata
HET/LRS   &  2011-11-01  &     600.00\\
UVOT/uvg  &  2011-11-02  &    3710.08\\
ARC/DIS   &  2011-11-11  &    1200.00\\
UVOT/uvg  &  2011-11-13  &  15,550.70\\
HET/LRS   &  2011-11-16  &     600.00\\
LBT/MODS1 &  2011-11-17  &     900.00\\
HET/LRS   &  2011-12-21  &     600.00\\
\enddata 
\end{deluxetable} 

\begin{deluxetable}{lcl} 
\tablecolumns{3} 
\tabletypesize{\scriptsize}
\tablecaption{SN~2011ht host galaxy properties
\label{tab3}} 
\tablewidth{0pt} 
\tablehead{
   \colhead{Parameter} &
    \colhead{Value} &
   \colhead{Note/Reference}
} 
\startdata 
Name   & UGC~5460 & NED  \\ 
R.~A.~(J2000)  & 10$^h$08$^m$09$\fs$3 & NED \\ 
Dec.~(J2000) & $+51$$\degr$50$\arcmin$38$\farcs$0 &  NED \\ 
Morphological Type & SB(rc)d & \citet{dvg91,chg94} \\ 
Heliocentric velocity & $1093$\,km\,s$^{-1}$ & NED \\ 
Virgo-infall corrected velocity & $1403$\,km\,s$^{-1}$ & NED \\ 
Distance modulus & 31.42\,mag & using $d_{\rm flow}=19.2$\,Mpc \\ 
$M_{B}$ & $-18.0$\,mag & this paper \\ 
Total SFR & $1.2^{+3.5}_{-1.0}$\,$\rm M_{\odot}$\,yr$^{-1}$ & this paper\\ 
Total M$_{\star}$  & $1.1^{+0.5}_{-0.4}\times 10^9$\,M$_{\odot}$ & this paper \\ 
Total A$_V$ & $0.8^{+0.6}_{-0.8}$\,mag & this paper \\ 
Oxygen abundance & $8.20$ & \citet{pp04}~$O3N2$ method, this paper \\ 
\enddata 
\end{deluxetable}

\begin{deluxetable}{crrcccr} 
\tablecolumns{7} 
\tabletypesize{\scriptsize} 
\tablecaption{SN~2011ht Blackbody and Bolometric Data 
\label{tab5}} 
\tablewidth{0pt} 
\tablehead{ 
  \colhead{UT} &
  \colhead{$T_{\rm{BB}}$} & 
  \colhead{$R_{\rm{BB}}$} & 
  \colhead{$L_{\rm{BB}}$} &
  \colhead{$L_{\rm{Pseudo}}$} &
  \colhead{$L_{\rm{Bolometric}}$} &
  \colhead{$\chi^2_\nu$} \\ 
  \colhead{(JD2450000+)} &
  \colhead{(K)} & 
  \colhead{($\times10^{14}$\,cm)} & 
  \colhead{($\times10^{42}$\,erg\,s$^{-1}$)} & 
  \colhead{($\times10^{42}$\,erg\,s$^{-1}$)} & 
  \colhead{($\times10^{42}$\,erg\,s$^{-1}$)} & 
  \colhead{ }  
}
\startdata
5838.56	&	4000	&	21.23	&	0.82	&	0.11	&	0.81	&	43.62	\\
5841.27	&	4800	&	13.23	&	0.66	&	0.16	&	0.66	&	22.96	\\
5842.21	&	5000	&	12.46	&	0.69	&	0.18	&	0.69	&	21.40	\\
5843.28	&	5300	&	11.24	&	0.71	&	0.22	&	0.71	&	22.76	\\
5852.53	&	9100	&	4.74	&	1.10	&	0.77	&	1.12	&	11.47	\\
5856.32	&	10100	&	4.59	&	1.56	&	1.18	&	1.61	&	3.03	\\
5858.69	&	10700	&	4.57	&	1.95	&	1.52	&	2.03	&	2.30	\\
5862.16	&	11800	&	4.40	&	2.68	&	2.15	&	2.79	&	2.05	\\
5865.40	&	12400	&	4.31	&	3.13	&	2.52	&	3.25	&	1.30	\\
5868.05	&	12700	&	4.33	&	3.48	&	2.80	&	3.61	&	2.61	\\
5871.05	&	13100	&	4.34	&	3.96	&	3.17	&	4.09	&	2.12	\\
5873.86	&	13200	&	4.46	&	4.29	&	3.43	&	4.43	&	2.30	\\
5876.84	&	13300	&	4.48	&	4.47	&	3.54	&	4.58	&	1.22	\\
5880.25	&	13200	&	4.53	&	4.43	&	3.54	&	4.57	&	1.23	\\
5883.23	&	13000	&	4.60	&	4.31	&	3.44	&	4.44	&	1.35	\\
5885.98	&	12900	&	4.62	&	4.22	&	3.38	&	4.35	&	2.85	\\
5889.03	&	12600	&	4.78	&	4.10	&	3.31	&	4.26	&	5.40	\\
5892.23	&	12100	&	5.06	&	3.91	&	3.15	&	4.07	&	2.08	\\
5895.01	&	11900	&	5.13	&	3.75	&	3.01	&	3.90	&	2.74	\\
5898.32	&	11200	&	5.40	&	3.26	&	2.60	&	3.40	&	3.42	\\
5901.33	&	10700	&	5.60	&	2.92	&	2.29	&	3.05	&	4.23	\\
5903.86	&	10400	&	5.82	&	2.82	&	2.20	&	2.95	&	3.36	\\
5907.41	&	9800	&	6.11	&	2.45	&	1.88	&	2.59	&	8.91	\\
5909.74	&	9400	&	6.38	&	2.27	&	1.71	&	2.41	&	7.24	\\
5913.12	&	8600	&	7.38	&	2.13	&	1.50	&	2.25	&	12.88	\\
5917.04	&	7600	&	9.11	&	1.97	&	1.23	&	2.08	&	20.37	\\
5920.25	&	7100	&	10.11	&	1.85	&	1.06	&	1.94	&	25.87	\\
5922.65	&	6700	&	11.22	&	1.81	&	0.94	&	1.87	&	27.09	\\
5930.27	&	5900	&	13.59	&	1.59	&	0.65	&	1.62	&	27.72	\\
5939.57	&	5100	&	18.09	&	1.58	&	0.46	&	1.58	&	29.41	\\
5944.19	&	4700	&	21.77	&	1.65	&	0.38	&	1.64	&	23.15	\\
5950.74	&	4300	&	25.23	&	1.55	&	0.27	&	1.54	&	22.78	\\
\enddata \end{deluxetable}

\begin{deluxetable}{crc} 
\tablecolumns{3} 
\tabletypesize{\scriptsize} 
\tablecaption{Blueshifted$^{{\rm a}}$ Line Measurements for SN~2011ht$^{{\rm b}}$ 
\label{tab4}} 
\tablewidth{0pt} 
\tablehead{ 
  \colhead{Spectral} & 
  \colhead{Emission} & 
  \colhead{Absorption}\\  
  \colhead{Line} & 
  \colhead{(km\,s$^{-1}$)} & 
  \colhead{(km\,s$^{-1}$)} 
}
\startdata
H$\alpha$     &  261  &  778\\
H$\beta$      &  211  &  747\\
H$\gamma$     &  194  &  706\\
H$\delta$     &  157  &  676\\
H$\epsilon$   &   96  &  670\\
H$\zeta$      &  129  &  646\\
H$\eta$       &  ---  &  646\\
              &       &     \\
He~I          &       &     \\
7065          &  218  &  ---\\
6678          &  145  &  558\\
5876          &  148  &  603\\
5015          &  106  &  519\\
4922          &   81  &  502\\
4471          &  217  &  563\\
4026          &  ---  &  609\\
3819          &  ---  &  606\\
              &       &     \\
O~I           &       &     \\
7774          &  ---  &  618\\
              &       &     \\
Ca~II~K       &  ---  &  643\\
\enddata 
\tablenotetext{a}{Relative to the heliocentric 
velocity of the galaxy.}
\tablenotetext{b}{These values correspond to the
MODS1 spectrum in Figure~\ref{fig-spect}.}
\end{deluxetable} 

\end{document}